\documentclass[iop]{emulateapj}

\usepackage{graphicx,epsfig,amssymb,layout,verbatim,rotating,calc,mathrsfs,epstopdf}
\usepackage[sumlimits,intlimits,namelimits]{amsmath}
\usepackage{graphics}
\usepackage{rotate}
\usepackage{enumerate}
\PassOptionsToPackage{linktocpage}{hyperref}
\usepackage{natbib}
\usepackage{footmisc}

\def\apjl{ApJL}

\def\apjs{ApJSS}
\def\be{\begin{equation}}
\def\ee{\end{equation}}

\def\lsim{\lower 2pt \hbox{$\, \buildrel {\scriptstyle <}\over
         {\scriptstyle \sim}\,$}}
\newcommand\gsim{\buildrel > \over \sim}
\begin{document}
\newcommand{\figureout}[3]{\psfig{figure=#1,width=5.5in,angle=#2}
   \figcaption{#3} }

\title{GAMMA-RAY EMISSION IN DISSIPATIVE PULSAR MAGNETOSPHERES:\\ FROM THEORY TO FERMI OBSERVATIONS}

\author{Constantinos Kalapotharakos\altaffilmark{1,2}, Alice K. Harding\altaffilmark{2} \&
Demosthenes Kazanas\altaffilmark{2}}

\affil{$^1$University of Maryland, College Park (UMDCP/CRESST), College Park, MD 20742, USA;\\
$^2$Astrophysics Science Division, NASA/Goddard Space Flight Center,
Greenbelt, MD 20771, USA}
\email{constantinos.kalapotharakos@nasa.gov}


\begin{abstract}
We compute the patterns of $\gamma$-ray emission due to curvature
radiation in dissipative pulsar magnetospheres. Our ultimate goal is
to construct macrophysical models that are able to reproduce the
observed $\gamma$-ray light-curve phenomenology recently published
in the Second \emph{Fermi} Pulsar Catalog. We apply specific forms
of Ohm's law on the open field lines using a broad range for the
macroscopic conductivity values that result in solutions ranging,
from near-vacuum to near Force-Free. Using these solutions, we
generate model $\gamma$-ray light curves by calculating realistic
trajectories and Lorentz factors of radiating particles, under the
influence of both the accelerating electric fields and curvature
radiation-reaction. We further constrain our models using the
observed dependence of the phase-lags between the radio and
$\gamma$-ray emission on the $\gamma$-ray peak-separation. We
perform a statistical comparison of our model radio-lag vs
peak-separation diagram and the one obtained for the \emph{Fermi}
standard pulsars. We find that for models of uniform conductivity
over the entire open magnetic field line region, agreement with
observations favors higher values of this parameter. We find,
however, significant improvement in fitting the data with models
that employ a hybrid form of conductivity; specifically, infinite
conductivity interior to the light-cylinder and high but finite
conductivity on the outside. In these models the $\gamma$-ray
emission is produced in regions near the equatorial current sheet
but modulated by the local physical properties. These models have
radio-lags near the observed values and statistically best reproduce
the observed light-curve phenomenology. Additionally, these models
produce GeV photon cut-off energies.
\end{abstract}


\pagebreak

\section{INTRODUCTION}

Almost half a century since the discovery of Pulsars
\citep{1968Natur.217..709H} the origin of their emission still
remains uncertain. This uncertainty derives both from our lack of
knowledge of the precise magnetospheric structure, the location of
the radiation emission region and also from the lack of knowledge of
the microphysical processes that produce the acceleration of the
radiating particles. However, progress has recently been achieved on
some of these issues.

For years the only known solution (analytic though) of the pulsar
magnetosphere was that of the retarded, inclined vacuum dipole
\citep[][hereafter, VRD]{1955AnAp...18....1D}. However, it was
apparent from the very beginning that this solution was unrealistic
because it produced huge electric fields with a component
$E_{\parallel}$ parallel to the magnetic field on the surface of the
star. Such fields would not only pull charges off the surface of the
star but would also initiate pair cascades that would fill the
magnetosphere with the number of charges necessary to short out
these fields everywhere in the magnetosphere
\citep{1969ApJ...157..869G}. The charge density needed to achieve
this is known as the Goldreich Julian density $\rho_{\rm GJ}$ {and
the solutions with $E_{\parallel}=0$ are known as Force-Free
Electrodynamic solutions (hereafter, FFE)}. These FFE magnetospheres
represent mathematically the simplest such structures, in the sense
that in the axisymmetric case they can be reduced to a single
equation for the poloidal magnetic flux, the so-called pulsar
equation \citep{1973ApJ...182..951S}. Even in this simplest of
cases, the structure of the magnetosphere remained unknown because
the pulsar equation is singular on the Light Cylinder (hereafter,
LC), a fact that stymied efforts to obtain numerical solutions.

The situation changed fifteen years ago when the first solution of
the pulsar equation that smoothly crossed the singular LC surface
was produced by \cite*{ckf99} (hereafter, CKF). This solution
determined, by numerical iteration, the poloidal current
distribution on the LC that allows a smooth transition of the
magnetic field lines across this surface.  In addition to the field
structure of the magnetosphere, this solution determined the global
current structure that flows through the entire magnetosphere and
also the local sign of its carriers. Because of the current flow
through the magnetosphere, any global solution must inherently
provide current closure. Thus, the CKF solution revealed the global
current flow of these magnetospheres: The current flows out mainly
on a current sheet along the last closed field surface (separatrix)
interior to the LC and the equatorial plane beyond the LC (there is
also some distributed current above this surface), while the return
current flows mainly along the dipole/rotational axis. The
axisymmetric solution has since been confirmed and further studied
by several others
\citep{2005PhRvL..94b1101G,timokhin2006,2006MNRAS.367...19K,
2006MNRAS.368L..30M,2003ApJ...598..446U,2011MNRAS.411.2461Y,2012MNRAS.423.1416P}.

The FFE magnetospheric solution set was completed when \cite{S2006},
using a time-dependent code, numerically integrated Maxwell's
equations appropriate for a rotating, oblique dipole (along with the
proper boundary conditions on the surface of the star) to produce
the first three-dimensional (hereafter, 3D) pulsar solutions. These
solutions reproduced the closed-open magnetosphere configuration of
CKF extending the solution to an oblique magnetic geometry.

\cite{kc2009} and \cite{2012MNRAS.424..605P} confirmed the structure
of these 3D solutions. The former used a time-dependent 3D scheme
similar to that of \cite{S2006} but with the incorporation of
Perfectly Matched Layer
\citep{1994JCoPh.114..185B,1996JCoPh.127..363B} at the outer
boundary of their computational domain, a technique that {minimizes
the} reflected waves at this boundary; this allowed them to follow
their simulations for many stellar periods out to distances $r
\simeq 10 \, R_{\rm LC}$ {(where $R_{\rm LC}$ is the LC radius)}.
The latter used a time-dependent pseudo-spectral 3D scheme where
they applied the Characteristic Compatibility Method in order to
avoid the inward reflection at their outer boundaries. {The main
features of the 3D ideal FFE solutions have also been confirmed by
recent relativistic magnetohydrodynamic models that inherently take
into account plasma inertia and pressure
\citep{2013MNRAS.435L...1T}}.

Similarly to the CKF, the 3D solutions also comprise an (undulating)
equatorial current sheet outside the LC that survives to large
distances \citep*{2012MNRAS.420.2793K}, while its shape is similar
to that of the split monopole solution current sheet
\citep{1999A&A...349.1017B}. However, within the LC, as the
inclination angle\footnote{The angle between the rotational axes and
the magnetic axes.} $\alpha$ increases, a progressively smaller
fraction of the return current reaches the surface of the star
through the current sheet. For $\alpha=90^{\circ}$ all the current
reaching the stellar surface is distributed over half of the polar
caps \citep{bs10b}. This means that, for $\alpha=90^{\circ}$, the
equatorial current sheet outside the LC is not directly connected to
the star.

In Fig.~\ref{fig01} we plot, in a color scale, the poloidal current
density $J_p$ distribution on the polar cap for the indicated
$\alpha$ values. The solid white line in all panels denotes the
polar cap rim. The open (closed) field lines start inside (outside)
this rim. {We note that any FFE simulation could have magnetic field
lines that close outside the LC due to unavoidable numerical
dissipation. Nonetheless, we define open magnetic field lines to be
those that cross the LC. In each panel there are notations that
indicate the directions (with respect to the magnetic axes) toward
the leading edge (LD), the rotational axis (AX), the trailing edge
(TR), and the rotational equator (EQ). This notation scheme is
followed in all the subsequent figures (whenever is needed).} The
positive values (reddish color) of $J_p$ indicate outward current
while the negative (blueish color) values indicate inward current.
The solid gray line is the zero current density line ($J_p=0$). We
see that for $\alpha=0^{\circ}$ (Fig.~\ref{fig01}a) there is a
current sheet all along the separatrix (the white rim). However, for
$\alpha>0$ (Figs.~\ref{fig01}b,c,d) only a part of the return
current reaches the {stellar surface in a current sheet form (the
areas within the dashed black lines). These areas are toward the
rotational axis and mostly toward the leading side of the polar cap,
becoming progressively smaller as $\alpha$ increases. This effect
makes, in general, asymmetric the structure from both sides of the
equatorial current sheet.

One should note here that both the analytic VRD and the numerical
FFE solutions are dissipationless and so, strictly speaking,
preclude the acceleration of particles and the emission of radiation
\citep{2012ApJ...749....2K}. The VRD solutions have huge
$E_{\parallel}$ but no particles ($\rho=0$) to accelerate, while the
FFE solutions have large values of the charge density $\rho$ which
however shorts-out the accelerating electric field $E_{\parallel}$
yielding again no acceleration. The observed, radiation-emitting
pulsars should therefore lie somewhere between these two limiting
regimes.

More recently, two groups
\citep{2012ApJ...749....2K,2012ApJ...746...60L} began filling the
solution gap between the VRD and FFE by producing solutions with
both $\rho \ne 0$ and $E_{\parallel} \ne 0$. In order to proceed
this way, one needs a macroscopic prescription for the current
density $J$, i.e. a form of Ohm's law. {The actual Ohm's law is
expected to be rather complicated incorporating many effects
\citep[e.g., pressure, inertia, Hall; see][]{2004ApJ...605..340M}
and taking into account also any pair production microphysical
processes. The adopted prescriptions, in the current study, even
though simplistic, provide solutions that have a distribution of
non-zero $E_{\parallel}$. All these prescriptions eventually involve
a conductivity parameter $\sigma$ (expressed in units of the
fundamental frequency in the problem, namely the pulsar rotation
frequency $\Omega$) that describes the plasma's ability to screen
the accelerating electric fields, with the final solution depending
on both the macroscopic prescription for $J$ and the corresponding
spatial distribution of $\sigma$.} According to most of these
prescriptions, as $\sigma$ goes from 0 to $\infty$, an entire
spectrum of solutions between VRD and FFE is recovered.

\begin{figure*}
  \centering
  \includegraphics[width=9cm]{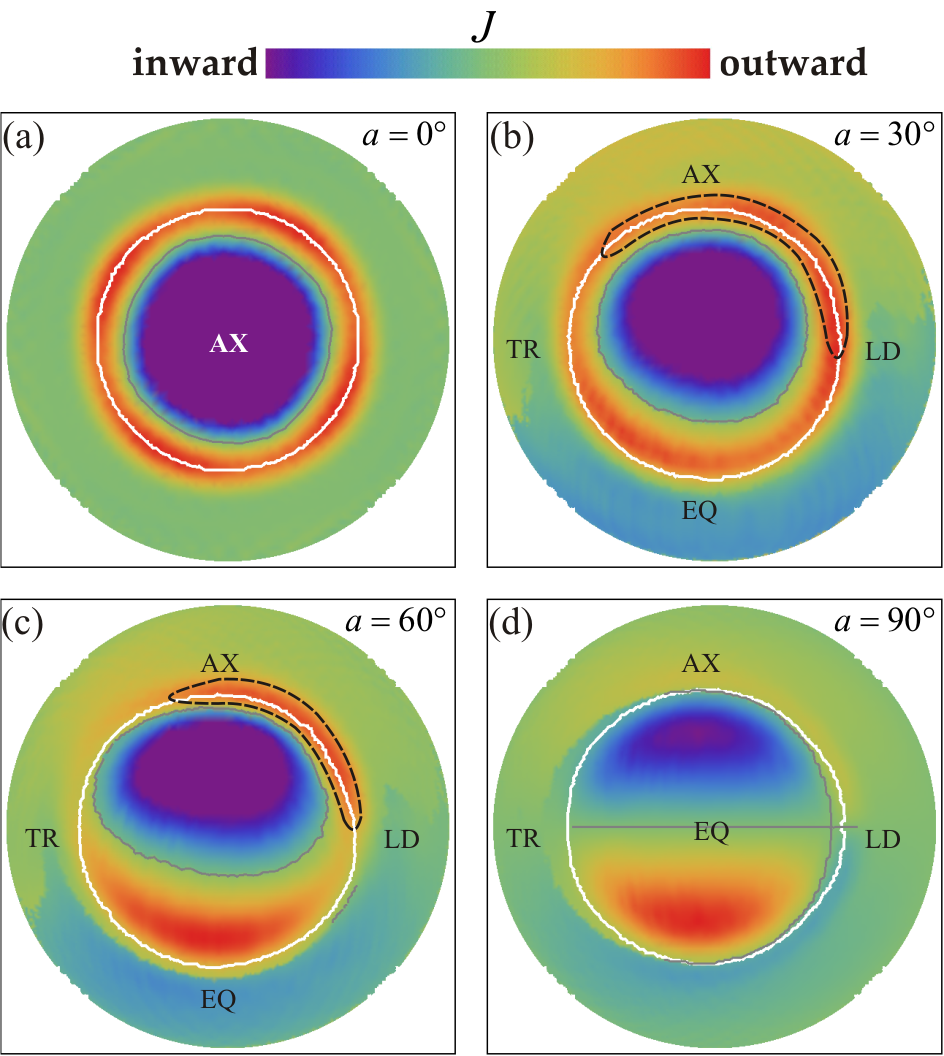}\\
  \caption{The distribution of the FFE poloidal current density
  on the polar caps (colors indicate the magnitude and direction) for the indicated
  inclination angles $\alpha$. In each panel the white line
  denotes the polar cap rim while the gray line denotes the zero
  current line that separates the inward from the outward
  current. For $\alpha=0^{\circ}$ the return current has
  a current sheet component all along the separatrix that
  separates the open from the closed magnetic field lines
  (strong red color along the white line). This current sheet
  component connects the equatorial current sheet with the
  stellar surface. As the inclination angle increases
  the return current that reaches the stellar surface in a
  current sheet form (part within the black dashed line)
  becomes gradually smaller until $\alpha=90^{\circ}$
  where the equatorial current sheet is not directly
  connected to the stellar surface. {The notations `LD', `AX', `TR',
  and `EQ' indicate the directions (with respect to the magnetic axis)
  toward the leading edge, rotational axis, trailing edge,
  and rotational equator, respectively. We note that the `AX' direction
  for $\alpha=0^{\circ}$ and `EQ' direction for $\alpha=90^{\circ}$
  coincide with the magnetic axis. The axisymmetric case ($\alpha=0^{\circ}$)
  is degenerate and the `LD', `TR', and `EQ' directions are not defined.}}\label{fig01}
\end{figure*}

On the observational side, the launch of {\em Fermi Gamma-Ray Space
Telescope} ushered in a new era in the study of pulsars. The
discovery of over 130 new pulsars to date in high energy ($> 100$
MeV) $\gamma$-rays \citep[117 of which are included in the Second
{\em Fermi} Pulsar Catalog (2PC), ][]{2013ApJS..208...17A},
increased the corresponding data base by twentyfold. The measurement
of exponential cut-offs in the pulsar spectra \citep{abdoetal2009}
immediately resolved the issue of the location of the high energy
$\gamma$-ray emission in favor of the outer magnetosphere.
Furthermore, the notable increase in the number of detections allows
meaningful statistical studies of their $\gamma$-ray properties and
cross-correlations with emission at other wavelengths, most notably
with the radio. One of the most important such correlations has been
that between the peak-separation $\Delta$ of the $\gamma$-ray pulses
and the lag $\delta$ between the $\gamma$-ray and radio emission.
This is important because it involves observed quantities across
widely separated frequency bands, one of which (radio) is widely
accepted as emitted by a region located near the pulsar dipole axis
from a region close to the pulsar surface. Since the $\gamma$-ray
emission depends more sensitively on the structure of the pulsar
magnetic field lines near the LC and $E_{\parallel}$ distribution,
the $\Delta - \delta$ diagram can be employed to test the pulsar
$\gamma$-ray emission models. The confrontation of pulsar model
magnetosphere predictions with the pulsar $\Delta - \delta$ diagram
is the essence of the present paper.

In \cite{2012ApJ...754L...1K} we presented for the first time
$\gamma$-ray light curves based on realistic particle trajectories
that take into account the physical properties (i.e. the field
structure and the values of $E_{\parallel}$) of dissipative
solutions. This study involved only one inclination angle
$\alpha=90^{\circ}$ but a wide range of conductivity values. The
resulting light curves exhibited double peaked shapes for solutions
corresponding to the entire range of $\sigma$ values. It was also
shown that as $\sigma$ increases, the corresponding emission region
moves outward and for the highest adopted $\sigma$ values (solutions
near-FFE) an important $\gamma$-ray component comes from a region
near the equatorial current sheet outside the LC. Nevertheless, all
these high-conductivity light-curves showed a clear trend for
radio-lags\footnote{The phase lag between the first $\gamma$-ray
peak and the radio-peak (thought to be near the magnetic pole, well
inside the LC).} larger than those observed. This ``lag'' problem
seems to appear also when the $\gamma$-ray emission is calculated by
employing simple phenomenological $\gamma$-ray emission models for
particles moving in FFE geometries
\citep{ck2010,bs10b,2011arXiv1111.0828H}.

In this paper, we extend our study of $\gamma$-ray light curves for
the entire range of values of the inclination angle $\alpha$. Our
goal is to study the radiation patterns of dissipative solutions and
to compare statistically the corresponding $\Delta - \delta$
diagrams with those observed. This will reveal whether and under
what assumptions the dissipative models can explain the observed
phenomenology. Eventually, the macrophysics of the models that
passes the comparison test successfully should be related to
microphysical mechanisms that can support it.

The structure of the present paper is as follows. In Section 2 we
give a brief outline of our dissipative models. In Section 3 we use
the field structure of these models to define realistic electron (or
positron) trajectories and calculate their corresponding energies
(including the effects of radiation losses) and the resulting
curvature radiation emission. In Section 4 we present our results.
We discuss the particle orbit properties, their evolution with
$\sigma$ and how this evolution determines the $\gamma$-ray emission
regions and the corresponding light curves. Particular emphasis is
given to the comparison between these results and the observational
data. In Section 5, guided by the results of Section 4, we present
special models corresponding to certain spatial $\sigma$
distributions that are able to reproduce the observed phenomenology.
Finally, in Section 6 we present our conclusions.

\section{Model Description}

In order to produce our models we solve numerically the time
dependent Maxwell equations
\begin{align}\label{maxeqb}
\frac{\partial \mathbf{B}}{\partial
t}&=-c\boldsymbol{\nabla}\times\mathbf{E}\\
\label{maxeqe} \frac{\partial \mathbf{E}}{\partial
t}&=c\boldsymbol{\nabla}\times\mathbf{B}-4\pi \mathbf{J}
\end{align}
using a 3D Finite Difference Time Domain technique
\citep{kc2009,2012ApJ...749....2K}. We consider the presence of a
dipole magnetic moment $\pmb{\mu}$ at the center of the star and
that the star itself is a perfect conductor; then the boundary
condition on the stellar surface for the electric field reads
$\mathbf{E}=-(\pmb{\Omega}\times\mathbf{r})\times\mathbf{B}/c$\,,
where $\pmb{\Omega}$ is the angular velocity of the neutron star and
$\mathbf{B}$ the magnetic field on its surface. The closure of the
system then requires a prescription for the current density
$\mathbf{J}$ in terms of the fields. For the FFE solutions this is
given by \citep{G1999}
\begin{equation}\label{IFFEJ}
    \mathbf{J}=c\rho\frac{\mathbf{E}\times\mathbf{B}}{B^2}+
    \frac{c}{4\pi}\frac{\mathbf{B\cdot \nabla\times B - E\cdot \nabla\times
    E}}{B^2}\mathbf{B}
\end{equation}
and guarantees that $\mathbf{E_{\parallel}}$, the component of the
electric field parallel to $\mathbf{B}$, be identically zero.
Numerically, rather than computing the above expression we employ
just the drift current (the first term in the above expression) and
we force the value of any {parallel} electric field component at
each time step to equal zero. One should note that the charge
density $\rho$ in the expression above is given by
$\rho=\pmb{\nabla}\cdot\mathbf{E}/(4\pi)$.

As noted above, realistic pulsar magnetospheres must be different
from those of FFE to allow for $\mathbf{E_{\parallel}} \neq 0$,
needed for the acceleration of particles that produce the observed
radiation and, therefore, one is compelled to produced non-FFE
models. This is achieved by considering prescriptions for
$\mathbf{J}$ different from those of Eq.~(\ref{IFFEJ}). In
\cite{2012ApJ...749....2K} we used the following very simplistic
prescription for the current density
\begin{equation}\label{ni1j}
    \mathbf{J}=c\rho\frac{\mathbf{E}\times\mathbf{B}}{B^2}+
    \sigma \mathbf{E_{\vert\vert}} \, .
\end{equation}
The first term is the usual drift component while the second term
regulates the $\mathbf{E_{\parallel}}$ through the conductivity
$\sigma$. This prescription requires a special treatment that sets
$E_{\perp}<B$ securing in that way that the drift component stays
always subluminal. This is most important near the equatorial
current sheet outside the LC where the magnetic field approaches 0.

We also produced models \citep{2012ApJ...749....2K} employing the
so-called Strong Field Electrodynamics \citep[][hereafter,
SFE]{2008arXiv0802.1716G} that has a covariant formulation for the
current density that reads
\begin{equation}\label{SFEJ}
    \mathbf{J}=\frac{c\rho \mathbf{E\times B}+(c^2\rho^2+
\gamma_f^2\sigma^2E_{0}^2)^{1/2}(B_0\mathbf{B}+E_0\mathbf{E})}{B^2+E_0^2}
\end{equation}
where
\begin{equation}\label{SFEINV}
    B_0^2-E_0^2=\mathbf{B^2-E^2},\; B_0E_0=\mathbf{E\cdot B},\; E_0\geq 0\end{equation}
\begin{equation}\label{SFEgam}
    \gamma_f^2=\frac{B^2+E_0^2}{B_0^2+E_0^2}~.
\end{equation}
This prescription as well as the one proposed by
\cite{2012ApJ...746...60L}
\begin{equation}\label{LSTJ}
    \mathbf{J}=\frac{c\rho \mathbf{E\times
    B}+\gamma_f\sigma E_0 (B_0\mathbf{B}+E_0\mathbf{E})}{B^2+E_0^2}~,
\end{equation}
where $E_0, B_0$ and $\gamma_f$ are defined by
Eqs.~\eqref{SFEINV},~\eqref{SFEgam}, have been produced considering
Ohm's law in the fluid frame. However, since this frame is not {\em
a priori} known, one has to assume a velocity. In SFE the
dissipation takes place only in the space-like regions ($J>\rho c$)
and the fluid frame is the one for which the charge density vanishes
$(\rho=0)$. In prescription \eqref{LSTJ} the fluid frame is assumed
to be the slowest moving one that has electric fields parallel to
magnetic fields.

All the above macroscopic prescriptions for the current density
$\mathbf{J}$ tend to the FFE prescription Eq. (\ref{IFFEJ}) as
$\sigma\rightarrow\infty$. Nevertheless, the advantage of
\eqref{ni1j} and \eqref{LSTJ} (though still not covariant) is that
their corresponding  models can span the entire spectrum of
solutions between VRD and FFE as $\sigma$ varies from 0 to $\infty$,
while SFE produces solutions that have $J \rightarrow \rho c$ as
$\sigma \rightarrow 0$. On the other hand, the SFE and \eqref{LSTJ}
naturally include the $E_0$ term in the denominator of the drift
term that makes the treatment of the region near the current sheets
much easier since it never allows the drift current to be
superluminal. Moreover, this expression is compatible with the
motion of the charged particles along the drift direction. Thus, in
the present work we decided to add the $E_0$ term in our
prescription \eqref{ni1j}, that now reads
\begin{equation}\label{ni1jR}
    \mathbf{J}=c\rho\frac{\mathbf{E}\times\mathbf{B}}{B^2+E_0^2}+
    \sigma \mathbf{E_{\vert\vert}}
\end{equation}
in order to take advantage of its features on the current
sheet and to be consistent to the motion of the particles along the
drift direction.

We note that the final solutions we get from prescriptions
\eqref{ni1j}, \eqref{LSTJ} and \eqref{ni1jR} do not differ
significantly\footnote{This is also valid for SFE in the high
$\sigma$ regime.} and one can always get the same spatial
distribution of $E_{\parallel}$ values just by applying a slightly
different value of $\sigma$. However, the distribution of $\sigma$
that determines the potential drops at each region of the
magnetosphere, which can reproduce the observed pulsar
phenomenology, the ultimate subject of this kind of research, cannot
be established beforehand.

\begin{figure}
  \centering
  \includegraphics[width=7.6cm]{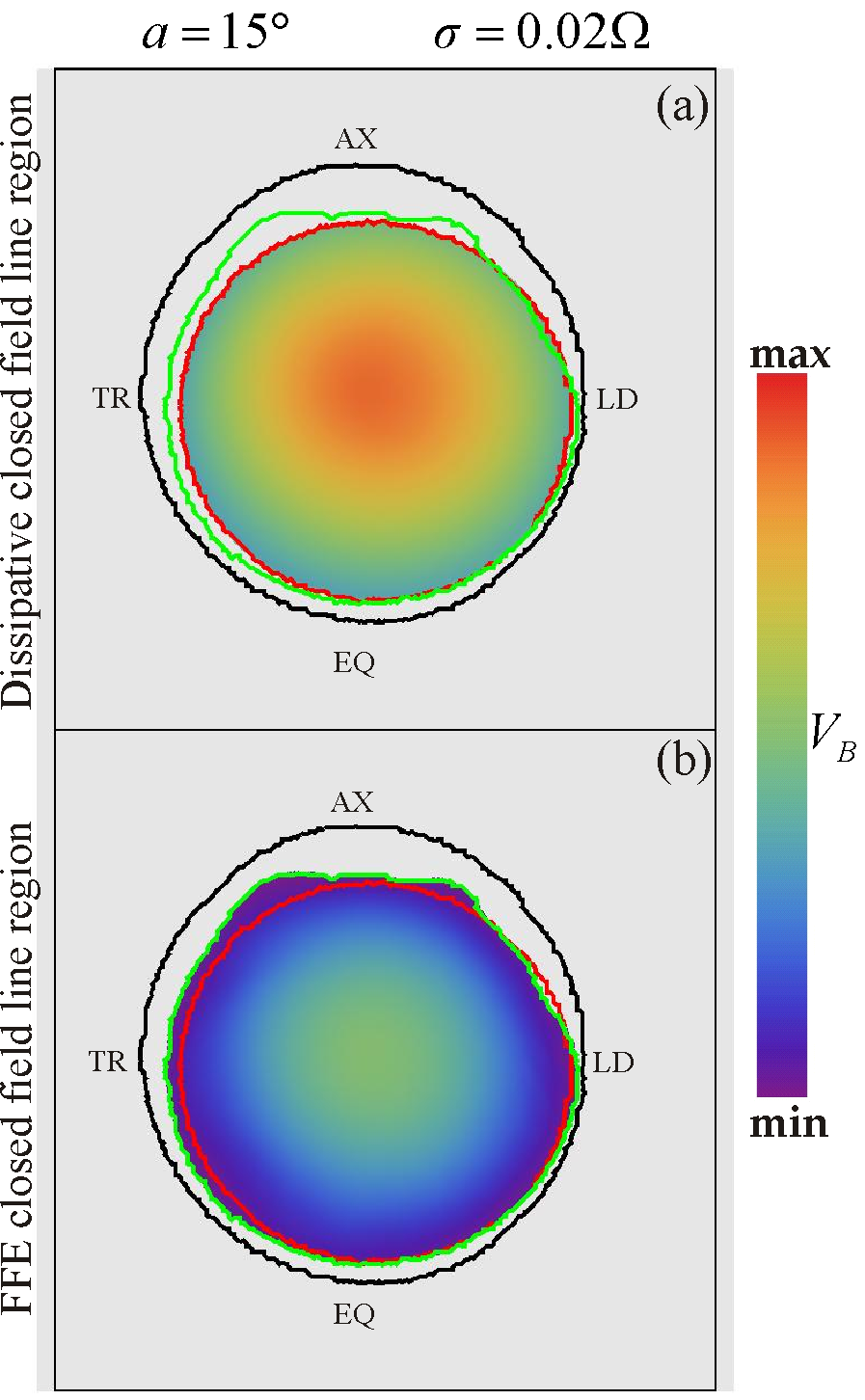}\\
  \caption{The distribution of the potential drop $V_{B}$
  along the magnetic field lines, in the indicated color scale,
  for $\alpha=15^{\circ}$. In both panels (a) and (b) the black line
  denotes the polar cap rim for the FFE solution. The red
  line denotes the polar cap rim for the dissipative
  solution that has $\sigma=0.02\Omega$ everywhere in the
  magnetosphere while the green line denotes the polar cap rim
  for the dissipative solution that has $\sigma=0.02\Omega$
  only in the open magnetic field line region and FFE regime for the
  closed region. {We see that the perfectly conductive closed region
  (b) decreases the potential drop along the open magnetic field lines
  compared to that of a magnetosphere that is dissipative in both
  open and closed regions (a).}
  We see also that the green polar cap rim is closer to the red one than it is
  to the FFE polar cap rim, even though
  it is larger. We note that for high $\alpha$ the corresponding
  green line is closer to the corresponding black line.}\label{fig02}
\end{figure}

In what follows, we present results produced employing prescription
\eqref{ni1jR} while we have tested that the main results remain
unaffected by the use of the other prescriptions. We have also
modified our code so that it can identify the different magnetic
field lines at each snapshot in time as the pulsar rotates. This
allows $\sigma$ to be magnetic-field-line dependent. This technique
will be extremely helpful when the microphysical studies will
provide all the different plasma properties along the different
magnetic field lines. However, this information is still incomplete
and so we reserve this kind of study for the future. Nevertheless,
we still employ this technique in the present paper in order to
produce solutions with different values of $\sigma$ in open and the
closed field lines. It is reasonable to consider that the closed
field line regions are described by FFE and only the `open' regions
may have small $\sigma$ values. Thus, for the solutions presented in
the next sections we have always applied very high $\sigma$ value
($\sigma=40\Omega$) for the closed field line regions. Note, that a
similar technique has been used by \cite{2012ApJ...746L..24L}. The
difference in that study is that it considers the closed field line
regions to be the same as the FFE ones, independently of the value
of $\sigma$ in the open field lines. However, this is not strictly
speaking correct, because one cannot determine the open-closed field
line boundary from a solution that has the same $\sigma$ (however
large) over all space, i.e. the FFE solution.
{\cite{2012ApJ...746L..24L} claimed that the spin-down, i.e. the
Poynting flux, is relatively insensitive to the size of the
conducting closed zone. However, for the current study the detailed
determination of the last open field lines and the corresponding
voltage along them is, in principle, essential.} This effect is more
dominant for low $\alpha$ values: In Fig.~\ref{fig02} we plot the
polar caps for $\alpha=15^{\circ}$ for $\sigma=0.02\Omega$ applied
everywhere in the magnetosphere (red line), for $\sigma=0.02\Omega$
applied only in the open field line region while the FFE {(in
reality highly conductive)} condition is applied in the `closed'
region (green line) and for the FFE {(everywhere applied)} solution
(black line). The color scale represents the potential drop $V_B$
along the magnetic field lines for the solution with finite
conductivity everywhere (Fig.~\ref{fig02}a) and for the solution
with finite conductivity only in the open field line region
(Fig.~\ref{fig02}b). We see that the green line is closer to the red
line (though still larger) rather than the black one. We note that
as we go to high $\alpha$ values the corresponding green line goes
closer to the black line.

Another issue is the construction of solutions of very high
$\sigma$. This problem arises from the small time steps $dt$ needed
to deal with the high $\sigma$ values of the `stiff' second term in
all expressions for the current density $\mathbf{J}$. The most
general solution of this problem is the use of implicit-explicit
integrators \citep[][and references therein]{2013MNRAS.431.1853P}.
However, in our case the solutions corresponding to
$\sigma\rightarrow\infty$ (FFE) are known. Near this regime (FFE),
the solutions have almost identical field structure and the only
difference is the spatial distribution of the small $E_{\parallel}$
values. The value $\mathbf{E_{\parallel}}$ is determined by the
following equation
\begin{equation}
\label{maxeqep} \frac{\partial \mathbf{E_{\parallel}}}{\partial
t}=c(\boldsymbol{\nabla}\times\mathbf{B})_{\parallel}-4\pi
\mathbf{J_{\parallel}}
\end{equation}
with  $\mathbf{J_{\parallel}}$ given by the second term of Eq.
(\ref{ni1jR}) and with
$(\boldsymbol{\nabla}\times\mathbf{B})_{\parallel}$ given by its FFE
value. For $\sigma\rightarrow\infty$, $E_{\parallel}\rightarrow 0$
and $\partial E_{\parallel}/(\sigma
\partial t) \rightarrow 0$. This implies that for high enough $\sigma$
the remaining (small) $E_{\parallel}$ can be calculated
approximately by setting the right hand side of Eq.~\eqref{maxeqep}
equal to 0. This $E_{\parallel}$ value can thus be obtained by the
expression
\begin{equation}
\label{eparapprox} E_{\parallel}=\frac{c
|(\boldsymbol{\nabla}\times\mathbf{B})_{\parallel}|_{\text{(FFE)}}}{4 \pi\sigma}
\end{equation}
where the $\boldsymbol{\nabla}\times\mathbf{B}$ is assumed to be the
one corresponding to FFE solutions. In case of prescriptions
\eqref{LSTJ} and \eqref{SFEJ} the $E_{\parallel}$ value can be
derived easily either numerically or from the linearized expressions
with respect to $E_{\parallel}$. The above approximation implies
that for a specific $\sigma$ value, the higher the
$(\boldsymbol{\nabla}\times\mathbf{B})_{\parallel}$ the higher the
$E_{\parallel}$ will be. {In order to confirm this approximation, we
ran a few simulations with
$\sigma=30\Omega~-~70\Omega~$\footnote{This was the highest $\sigma$
value we could handle using a grid size of 0.005$R_{\rm LC}$ and a
computational domain of $\approx 1800^3$ cells.} using prescriptions
\eqref{ni1jR} and \eqref{LSTJ} and we compared these models to the
corresponding FFE ones that have the approximated spatial
$E_{\parallel}$ distributions (e.g. Eq.~\ref{eparapprox}).} There
were only slight differences between these two configurations that
do not affect the trajectories and the corresponding light-curve
results we present in the next sections.

In Fig.~\ref{fig03} we plot, in color scale, and for the indicated
$\alpha$ values, the distribution of $E_{\parallel}$ provided by
relation \eqref{eparapprox}. In this plot we have considered the
same $\sigma$ value (e.g. $\sigma=500\Omega$) everywhere. The blue
(red) color implies $E_{\parallel}$ antiparallel (parallel) to the
local $B$. The distribution of $E_{\parallel}$ reflects the
distribution of $(\boldsymbol{\nabla}\times\mathbf{B})_{\parallel}$
in the corresponding FFE solutions. {For the constant $\sigma$ (in
the open field zone) solutions, in the remaining of this study we
employ approximation \eqref{eparapprox} for all cases with
$\sigma\geq30\Omega$.}

In our simulations we use a stellar radius $r_{\star}=0.2R_{\rm
LC}$\footnote{In our plots the length unit is always the $R_{\rm
LC}$.}. Our computational domains extend up to $5R_{\rm LC}$ and the
spatial resolution (i.e. the grid cell size) is $0.01R_{\rm LC}$.
Each simulation has been evolved for 2-3 stellar rotations. {We note
that the calculations presented below assume standard values for
pulsar period $P_{\star}=0.1s$ and magnetic field on stellar surface
$B_{\star}=10^{12}G$.}

\begin{figure*}
  \centering
  \includegraphics[width=\textwidth]{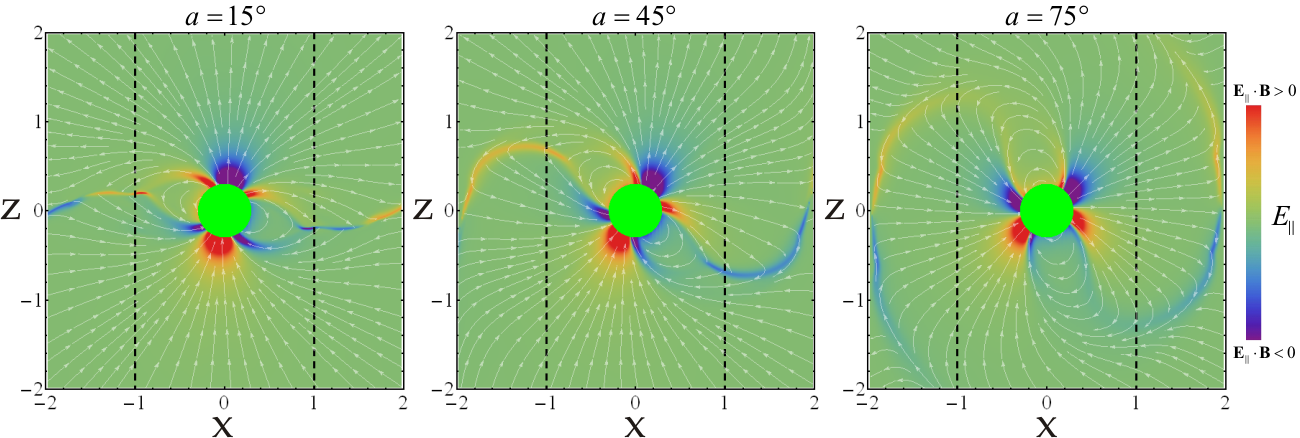}\\
  \caption{The distribution of $E_{\parallel}$, in the indicated
  color scale, on the poloidal plane $(\pmb{\mu},~\pmb{\Omega})$,
  for very high conductivity values (near-FFE) and for the indicated
  $\alpha$ values. We see, in general, high $E_{\parallel}$ above the
  polar caps, near the separatricies, and near the equatorial
  current sheets outside the light-cylinder (vertical black dashed lines).}\label{fig03}
\end{figure*}

\section{Realistic Particle Trajectories}

In the present paper we study the contribution of curvature
radiation to the observed $\gamma$-ray light curves. We define
realistic trajectories of charged particles ($e^-$ and $e^+$)
following a methodology similar to that presented in
\cite{2012ApJ...754L...1K}. Assuming that the speed of the particles
is always very close to $c$, we set the velocity vector
$\mathbf{v}$, everywhere in the magnetosphere, to be
\begin{equation}
\label{velocityv}
\mathbf{v}=\left(\frac{\mathbf{E}\times\mathbf{B}}{B^2+E_0^2}+f
\frac{\mathbf{B}}{B}\right)c~.
\end{equation}
The first term in expression \eqref{velocityv} is a drift velocity
component similar to that in our $\mathbf{J}$ prescriptions while
the second term is a component along the local magnetic field line.
The sign and the value of the scalar quantity $f$ {is set so} that
the total speed of the particle be $c$ and the motion be outward.
For a given magnetospheric solution (i.e. specific field
configuration) the spatial distribution of $f$ is then uniquely
determined. This fixes the assumed velocity flow in the entire
magnetosphere and hence the particle trajectories. Since the
particle trajectories are defined, the only additional requirement
for the calculation of the particles' kinetic energy along their
trajectories is their initial conditions (position and energy).
Assuming we know these we can calculate the $\gamma_L$ values
(Lorentz factors) along each trajectory taking into account their
acceleration by the $\mathbf{E_{\parallel}}$ provided by the
solution and their losses due to curvature radiation. Thus, the
calculation of the $\gamma_L$ value is made by integration over time
of
\begin{equation}
\label{dgdt} \frac{d\gamma_L}{dt}=f\frac{q_e c E_{\parallel}}{m_e
c^2}-\frac{2q_e^2\gamma_L^4}{3R_{\rm CR}^2m_e c}
\end{equation}
where $q_e$ and $m_e$ are the electron charge and rest-mass,
respectively while the quantity $R_{\rm CR}$ is the radius of
curvature at each point of the trajectory. The first term in
\eqref{dgdt} formulates the energy gain rates of the particles due
to the parallel electric field component they encounter along their
trajectories and the second term formulates the energy loss rates
due to curvature radiation.

Below, whenever it is not stated otherwise, the radiating particles
are considered to start on the stellar surface, distributed
uniformly on the polar cap with small $\gamma_L$ values
($\gamma_L<100$). {In each model, we integrated $\approx
2\times10^6$ trajectories originating on each polar cap. We have
checked that this number ensures a reliable statistics and provides
robust results. We note, also, that at each point of the stellar
surface we assume that we have the kind of particles ($e^-, e^+$)
needed in the region to accelerate outward.} We integrate these
trajectories up to $r=2.5R_{\rm LC}$, we collect the bolometric
energy along the emitting directions (locally tangential to the
trajectories) and taking into account the time-delay effects, we
construct sky-maps and the corresponding light-curves.

\begin{figure*}
  \centering
  \includegraphics[width=11.5cm]{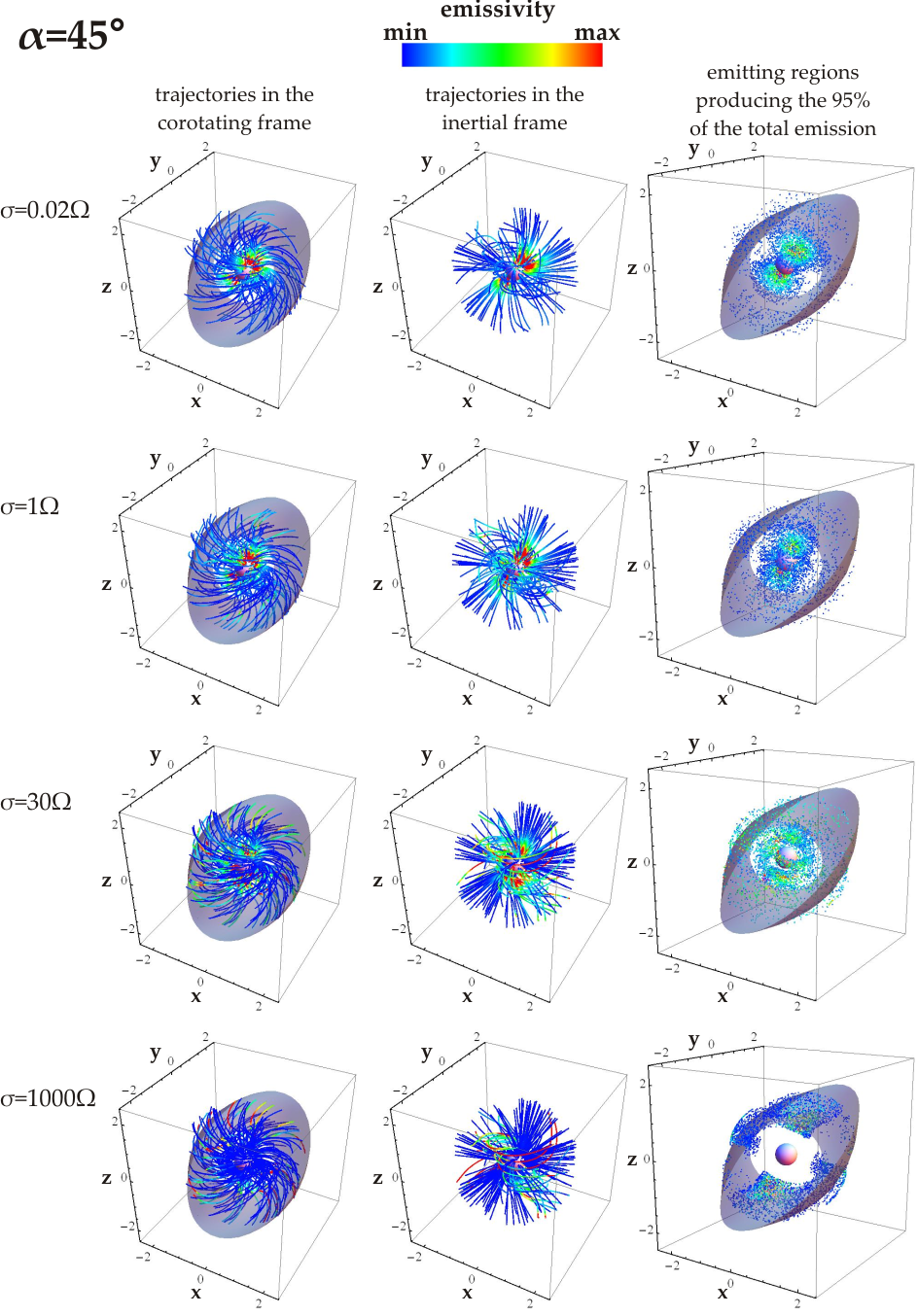}\\
  \caption{Trajectories and emission level (in the indicated color scale)
  for $\alpha=45^{\circ}$ and for different $\sigma$ values.
  \textbf{(Left hand column)} We plot the particle trajectories
  in the corotating frame together with the equatorial current sheet.
  The color along the orbits represents the local emissivity due to
  curvature radiation $\propto\gamma^4 R_{\rm CR}^{-2}$.
  \textbf{(Middle column)} The same trajectories but in the
  inertial frame. \textbf{(Right hand column)} {A sampling of points
  that trace the comoving (prescribed by the particle flux) magnetosphere volume
  contributing to the highest 95\% of the total emission. We note that each
  of these points represent the same number of particles and not equal volume elements.}
  All columns show that as $\sigma$ increases the emission
  moves gradually outward and for high $\sigma$ values it is produced in
  regions near the equatorial current sheet outside the light-cylinder.
  We note also that the trajectories for $\sigma\geq30\Omega$ are almost
  identical since for these $\sigma$ values the field structure remains
  almost the same and the particles are near corotation along the magnetic
  field lines.}\label{fig04}
\end{figure*}

\section{Results}
\label{sec_results}

\subsection{Radiation patterns}

In the left-hand column of Fig.~\ref{fig04} we plot, for
$\alpha=45^{\circ}$ and for the indicated $\sigma$ values, a sample
of 200 particle trajectories in the corotating frame. The gray
transparent surface represents the equatorial current sheet outside
the LC as it is defined in the corresponding FFE solution. The color
along each trajectory represents the local emissivity ($\propto
\gamma_L^4 R_{\rm CR}^{-2}$) according to the indicated color scale.
In the middle column, the same trajectories are plotted but in the
inertial frame. We note that the trajectories, for high $\sigma$
follow very closely the magnetic field lines in the corotating frame
while the trajectories of low $\sigma$ deviate from them since the
first term of the right-hand side of Eq.~\eqref{velocityv} {no
longer supports corotation.} For low $\sigma$ we see significant
radiation (reddish color) at low altitudes, well within the LC.
However, as $\sigma$ increases radiation starts being produced at
higher altitudes. Thus, the higher the value of $\sigma$, the higher
is the altitude along each trajectory  at which most of the emission
is produced. This happens because as $\sigma$ increases
$E_{\parallel}$ decreases and so the particles need to traverse
longer distances to reach the $\gamma_L$ values that allow them to
radiate efficiently. Moreover, the trajectories that produce
significant emission in the outer parts of the magnetosphere seem to
be those moving in regions near the equatorial current sheet in the
corotating frame (see left-hand column of Fig.~\ref{fig04}). {In the
right-hand column we plot, in the corotating frame, the points that
trace the comoving\footnote{This comoving volume element varies
inversely proportional to the particle number density with the
corresponding emission being simply the emission per unit mass
(rather than per unit volume).} volume (prescribed by the particle
flux) where the highest 95\% of the total emission of the entire
magnetosphere is produced.} The color of individual points still
represents the corresponding emissivity. These figures show that for
low $\sigma$ almost all the emission comes from the inner
magnetosphere from lobes above the magnetic poles. For higher
$\sigma$ these lobes change their orientation slightly and become
progressively smaller with $\sigma$. At the same time more and more
points from the outer magnetosphere contribute to the total
emission. These (outer) points, for high $\sigma$ ($\sigma>1\Omega$)
lie in regions close to the equatorial current sheet. Actually, for
very high $\sigma$ values the inner magnetosphere {has no
significantly radiating points} and all the emission is produced in
the outer magnetosphere in regions near the equatorial current
sheet.

\begin{figure*}
  \centering
  \includegraphics[width=12cm]{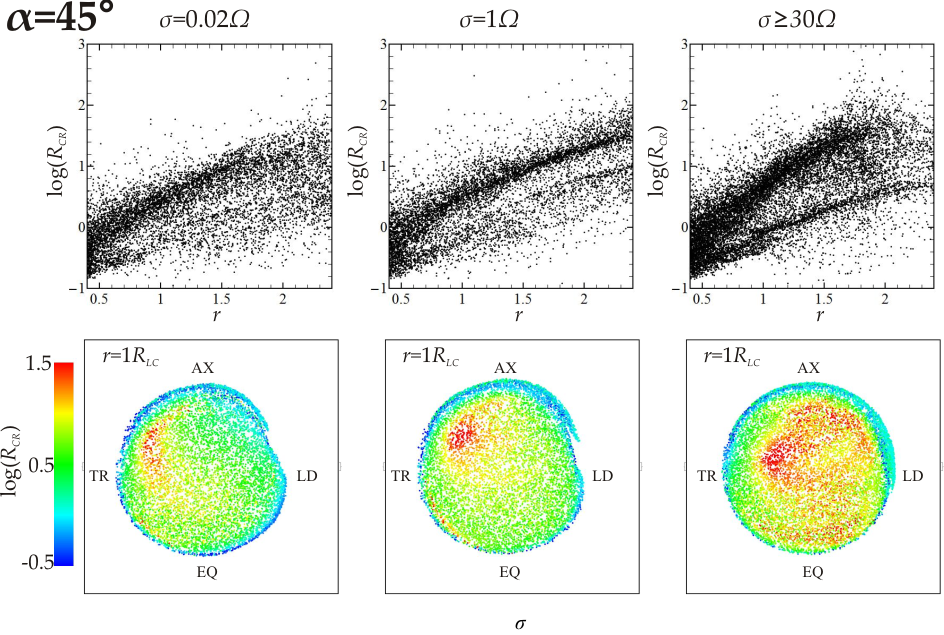}\\
  \caption{The radius of curvature $(R_{\rm CR})$ measured in light-cylinder
  radii for $\alpha=45^{\circ}$ and for the indicated $\sigma$ values.
  \textbf{(Top row)} We plot the $\log(R_{\rm CR})$ vs spherical radius
  $r$ {(measured in light-cylinder radii)} along many particle trajectories
  that start on the corresponding polar caps. We see that
  $R_{\rm CR}\propto 10^r$ with a scatter of one order of magnitude.
  \textbf{(Bottom row)} The $\log(R_{\rm CR})$ values
  of particle trajectories at $r=1R_{\rm LC}$ mapped (in the indicated color scale)
  on the corresponding polar caps. We see that particles trajectories that
  start near the polar cap edges have smaller $R_{\rm CR}$ values.}\label{fig05}
\end{figure*}

The local emissivity is more sensitive to the local values of
$\gamma_L$ than the corresponding values of $R_{\rm CR}$. In turn,
the local values of $\gamma_L$ are very sensitive to the values of
$E_{\parallel}$ that the particles see up to a specific point, while
$R_{\rm CR}$ is not as sensitive to $E_{\parallel}$ but depends
mostly on the geometry of the magnetic field lines. Actually,
$R_{\rm CR}$ remains almost unchanged beyond some not extremely high
value of $\sigma$. In the first row of Fig.~\ref{fig05} we plot, for
$\alpha=45^{\circ}$ and for the indicated values of $\sigma$, the
values of $\log(R_{\rm CR})$ versus the spherical radius $r$ for
many points along a large sample of particle trajectories. We see
that, in general, $\log(R_{\rm CR})$ increases almost linearly with
$r$, even though there is a dispersion of $\log(R_{\rm CR})$ values
that scatters the corresponding $R_{\rm CR}$ values within 1-2
orders of magnitude. The radius of curvature can be approximated by
the expression $R_{\rm CR}\propto 10^r$. For a specific $r$ the
larger $R_{\rm CR}$ values correspond to the inner region of the
polar cap while the smaller ones come from trajectories originating
near the polar cap rim (last open field lines). In the second row of
Fig.~\ref{fig05} we map on the polar cap, in the indicated color
scale, the values of $\log(R_{\rm CR})$ that these trajectories
reach at $r=1R_{\rm LC}$. These plots show clearly that $R_{\rm CR}$
increases slightly with $\sigma$. Moreover, we see that the small
$R_{\rm CR}$ values are always for trajectories that originate near
the polar cap edge. This can be also observed in the middle column
of Fig.~\ref{fig04}, where it is clear that the trajectories that
start near the edge of the polar cap are much more curved (i.e.
smaller $R_{\rm CR}$) than the trajectories originating inside the
polar cap rim. This implies that the high emission observed in
regions near the equatorial current sheet, in the high $\sigma$
models is additionally enhanced by the smaller values of $R_{\rm
CR}$ of these trajectories. {The general properties and trends
presented in Figs.~\ref{fig04}, \ref{fig05} remain the same for all
values of the pulsar inclination angle $\alpha$. The $R_{\rm CR}$
values and their dependence on $r$, solely determined by the field
structure, put certain restrictions on the location where specific
photon energies could be produced. The cut-off energy in the
curvature radiation spectrum is given by
\begin{equation}
\label{cutoff} \epsilon_c=\frac{3}{2}c \hbar\frac{\gamma_L^3}{R_{\rm
CR}}.
\end{equation}
Figure~\ref{fig06} is a contour plot of $\log(\epsilon_c)$ in the $r
- \log(\gamma_L)$ plane. In this plot we have considered that
$R_{\rm CR}=0.05\times10^r$ (an approximate expression that
describes the lower segment of the points plotted in
Fig.~\ref{fig05} for $\sigma\geq 30\Omega$). The two thick black
lines denote the indicated $\epsilon_c$ values (1MeV and 1GeV) while
the horizontal dashed white line denotes the $r=1R_{\rm LC}$ value.
Figure~\ref{fig06} shows the minimum $\gamma_L$ value required for a
specific photon energy to be emitted at a particular distance. For
example, in order to have GeV photon emission at $r = 1R_{\rm LC}$
the emitting particles must have Lorentz factors at least
$\gamma_L=10^{7.3}$ (at this distance).}

\begin{figure}
  \centering
  \includegraphics[width=6cm]{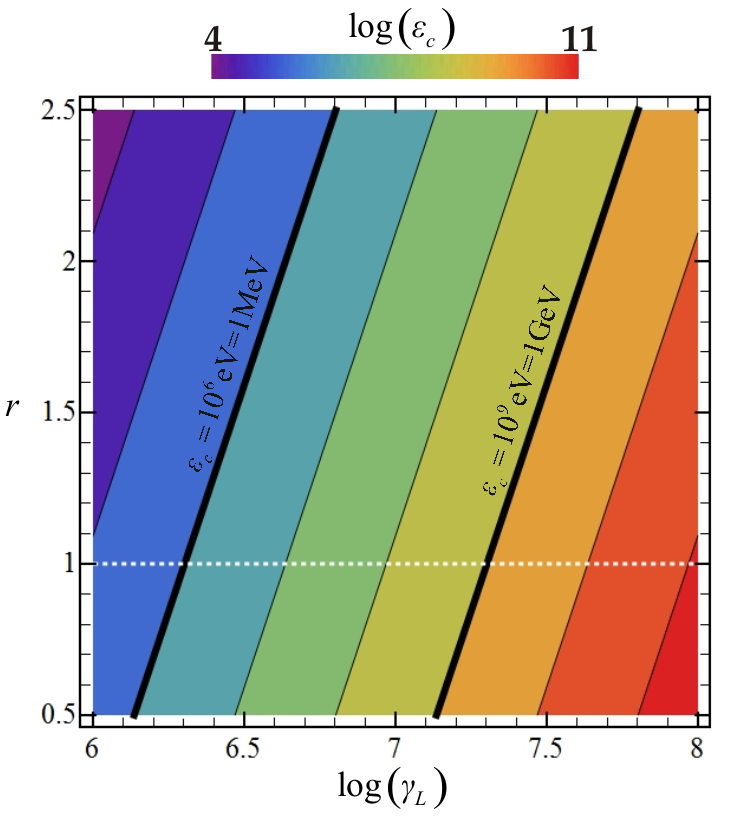}\\
  \caption{{The cut-off energy $\epsilon_c$ in the indicated logarithmic scale
  as a function of $\log(\gamma_L)$ and $r$. The two thick black lines denote
  the 1Mev and 1GeV $\epsilon_c$ values, as indicated in the figure.
  The white dashed horizontal line denotes
  the $r=1R_{\rm LC}$ value. There are restrictions with respect to the
  minimum $\gamma_L$ values required for certain photon energy emission
  at some distance.}}\label{fig06}
\end{figure}

\begin{figure*}
  \centering
  \includegraphics[width=12cm]{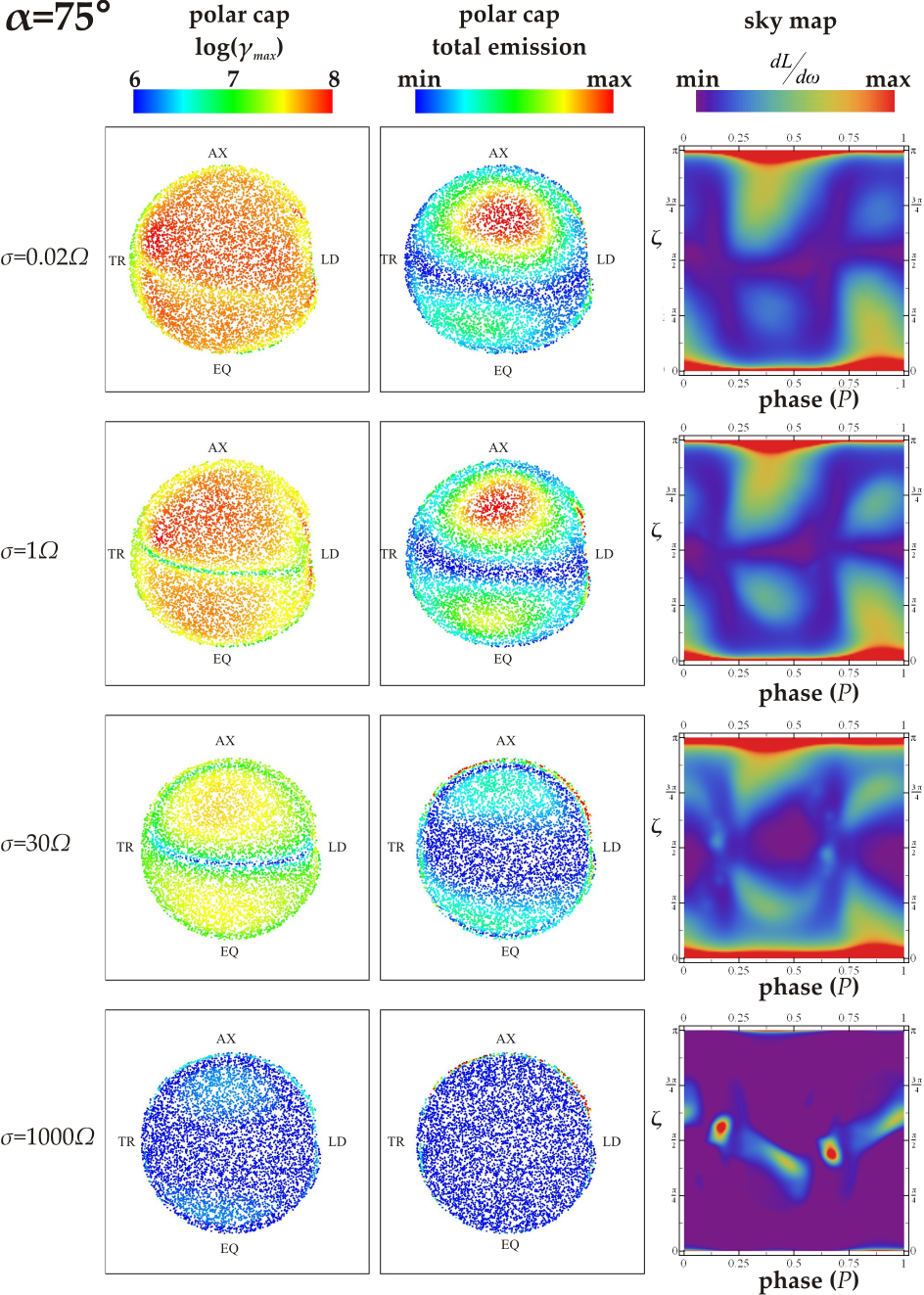}\\
  \caption{Radiation patterns and $\gamma_L$ values for $\alpha=75^{\circ}$
  and for the indicated $\sigma$ values. \textbf{(Left hand column)}
  The maximum $\gamma_L$ values of particle trajectories up to
  $r=2.5R_{\rm LC}$ mapped, in logarithmic color scale, on the
  corresponding polar caps. For low $\sigma$ values the $\gamma_{L_{max}}$
  values can reach higher than $10^8$ while for very high $\sigma$ values
  hardly exceed $10^6$. \textbf{(Middle column)} The total emission values
  along the particle trajectories up to $r=2.5R_{\rm LC}$ mapped,
  in the indicated color scale, on the corresponding polar caps. We see that the high
  total emission values move from the inner part of the polar cap toward
  its edge. {Note that in these panels we plot a sampling of points
  $(\approx 10^4)$ that is 1/200 of the total number $(2\times 10^6)$ of
  particle trajectories we have integrated in each polar cap.}
  \textbf{(Right hand column)} {The corresponding sky-maps: the luminosity per
  solid angle $dL/d\omega$, in the indicated color scale as a function of the rotation
  phase $\phi_{ph}$ and the observer inclination angle $\zeta$.}}\label{fig07}
\end{figure*}

In Fig.~\ref{fig07} we show in color scale the polar cap maps of the
$\log$ of the maximum value of $\gamma_L$ along each trajectory
($\gamma_{L_{\max}}$, left-hand column) and the total (along the
entire trajectory) emission (middle column) for $\alpha=75^{\circ}$
and the indicated $\sigma$ values. We see that for low $\sigma$
models the particles can reach values up to
$\gamma_{L_{\max}}\gtrsim 10^8$. These values are reached in the
inner magnetosphere (well inside the LC) with their upper limiting
values determined by the radiation reaction limit (curvature
radiation energy loss rate equal to energy gain due to
$E_{\parallel}$). The values of $\gamma_{L_{\max}}$ decrease with
increasing $\sigma$ and for $\sigma=1000\Omega$ most of them drop
below $10^6$ with only the highest ones reaching the value $\simeq
10^{6.5}$.

The total emission map (middle column) shows a similar pattern but
it nevertheless differs since it is modulated by the $R_{\rm CR}$
values and the distribution of the $\gamma_L$ values all along the
trajectories. The patterns shown in the left-hand and middle columns
of Fig.~\ref{fig07} resemble that of the current density
distribution on the polar cap (see Fig.~\ref{fig01}) confirming,
generally, that $E_{\parallel}$ is higher wherever the poloidal
current is higher. However, as $\sigma$ increases and
$E_{\parallel}$ decreases, $\gamma_L$ becomes sufficiently large to
emit significant curvature radiation at increasingly larger
distances. In these high $\sigma$ cases, the most efficiently
radiating trajectories are those that see high $E_{\parallel}$ at
large distances (beyond the LC). These trajectories are those
reaching close to the equatorial current sheet near which we
generally have high $E_{\parallel}$ {{(because the current density
requirement in the FFE regime in this region is high)}}. These
trajectories originate near the edge of the polar cap (last open
field lines) and that is why we observe (Fig.~\ref{fig07}) in these
regions, gradually, relatively higher $\gamma_{L_{\max}}$ and total
emissions (than in the other regions). Nevertheless,
Fig.~\ref{fig07} shows clearly that the total emission is not
uniformly distributed over the region near the equatorial current
sheet (i.e. the emission is not uniformly distributed all over the
polar cap rim).

\begin{figure*}
  \centering
  \includegraphics[width=\textwidth]{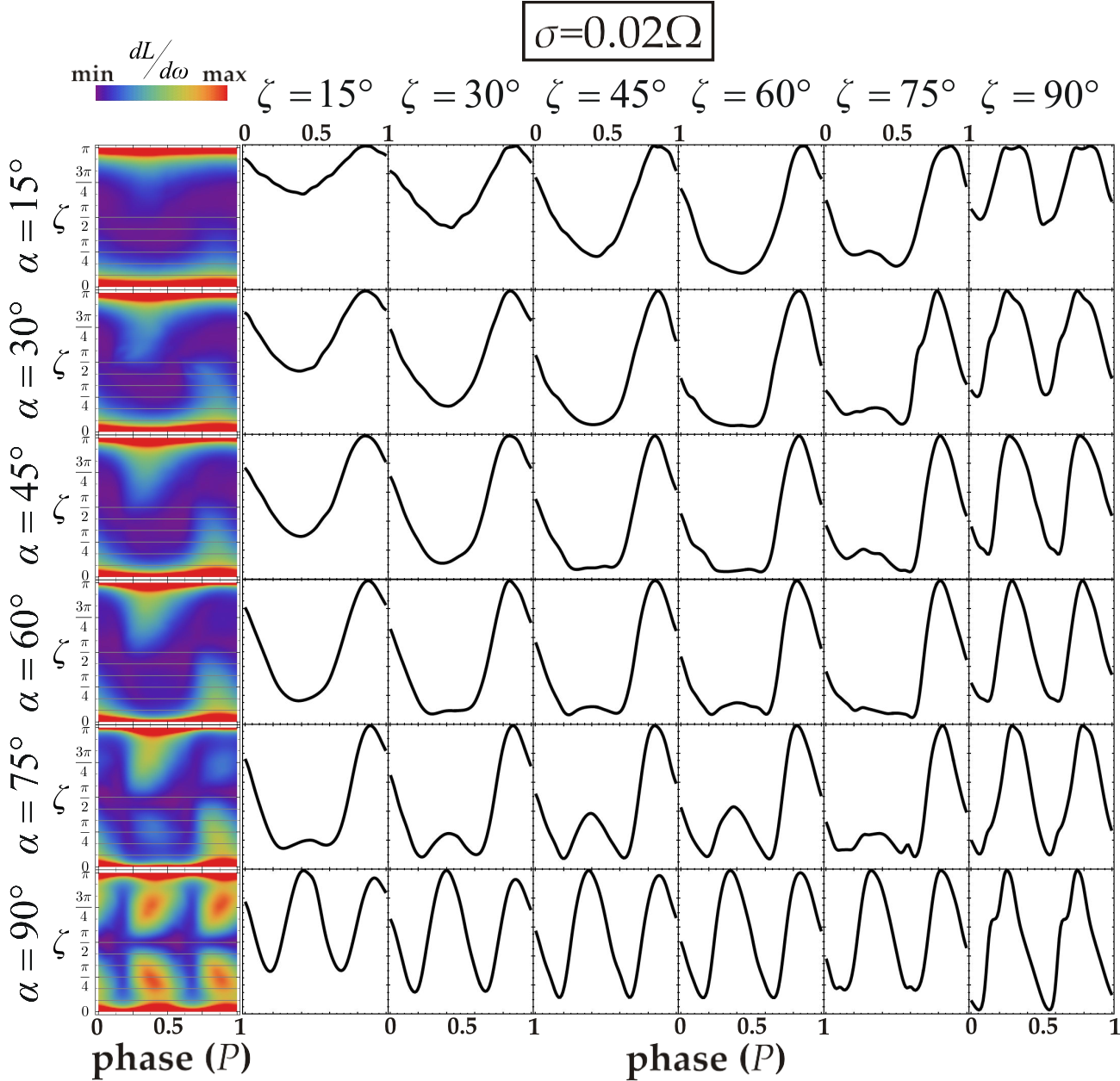}\\
  \caption{The $\gamma$-ray light-curve atlas together with the corresponding
  sky-maps for $\sigma=0.02\Omega$. In each sky-map, for different inclination
  angles $\alpha$, gray
  horizontal lines are plotted at the $\zeta$ values for which the light-curves
  are displayed. In this case all the emission comes from the inner
  magnetosphere within the light-cylinder. The corresponding light curves are not
  very narrow.}\label{fig08}
\end{figure*}

As we mentioned above, for high $\sigma$ the values of
$E_{\parallel}$ are small and consequently the particle energies and
the corresponding $\gamma_L$ values are not efficient in producing
significant emission in the inner magnetosphere (within the LC).
However, the $\gamma_L$ values with which the particles reach the LC
are not equal. The distribution of the $\gamma_L$ values on the LC
depends on what $E_{\parallel}$ the particles encounter up to the
LC. For small $E_{\parallel}$ (i.e. high $\sigma$) the $\gamma_L$
value depends mostly on the energy gain (first term in the
right-hand side of Eq.~\ref{dgdt}) up to the LC (since the
corresponding energy loss due to curvature radiation is small). This
essentially means that the particles starting on the polar cap
regions where the current reaches the stellar surface in a current
sheet form (see Fig.~\ref{fig01}) will have the highest $\gamma_L$
values at the LC because they encounter (amongst all polar cap
particles) the highest $E_{\parallel}$ values within the LC. The
non-uniformity of the $\gamma_L$ values on the LC affects the
relative efficiency of the radiation beyond this point due to the
$E_{\parallel}$ these particles see in the regions near the
equatorial current sheet outside the LC. Higher $\gamma_L$ value on
the LC means stronger emission beyond the LC near the equatorial
current sheet.

The previous study provides the direction of the emitted photons
(tangentially to the trajectories) and the corresponding bolometric
emission ($\propto \gamma_L^4 R_{\rm CR}^{-2}$) along each
trajectory. This information allows us to construct $\gamma$-ray
light curves by collecting all the photons in sky-maps taking into
account time-delay effects. {In the right-hand column of
Fig.~\ref{fig07} we plot the sky-maps for $\alpha=75^{\circ}$ and
for the indicated $\sigma$ values. Each sky-map depicts, in the
indicated color scale, the emitted luminosity per solid angle
($dL/d\omega=dL/\sin(\zeta)d\zeta d\phi_{ph}$). The horizontal axis
represents the phase of the pulsar rotation while the vertical axis
represents the observer inclination angle\footnote{This is the angle
between the rotational axis and the line of sight.} $\zeta$. The
zero phase is assigned to photons emitted close to the stellar
surface on the $\boldsymbol{\mu} {\bf - \Omega} $ plane.} Each
observer sees a different light curve depending on its $\zeta$
value. The observed light curve is the horizontal cut of the sky-map
at the observer's $\zeta$ value.

The high emission we observe for low $\sigma$ values
(Fig.~\ref{fig07}, right-hand column for $\sigma=0.02\Omega$,
$\sigma=1\Omega$) comes from lobes of points above the polar caps
well within the LC. As $\sigma$ increases these lobes become
smaller, making the emitting regions on the sky-maps narrower while
gradually there appear additional components coming from regions
near the equatorial current sheet. For very high $\sigma$
($\gtrsim300\Omega$) all the emission comes from specific regions
near the equatorial current sheet (see bottom panel in middle-column
of Fig.~\ref{fig07}) and the corresponding emitting regions on
sky-maps are narrow.

The sky-map pattern shown in Fig.~\ref{fig07} for
$\sigma=1000\Omega$ is quite different than that corresponding to
uniform emission in the entire equatorial current sheet. As we
explained above, for very high $\sigma$ values, the trajectories
that start from polar cap regions, where the current reaches in a
current sheet form, radiate much more effectively beyond the LC.
However, for the smaller $\alpha$ values ($\leq 45^{\circ}$) these
regions on the polar cap are quite extended and so the corresponding
sky-maps are very similar to those produced by uniform emission over
the entire equatorial current sheet. In the bottom panel of
right-hand column of Fig.~\ref{fig04} we see that emission is
produced by the largest part of the equatorial current sheet (though
still {not} totally uniform). On the other hand for larger $\alpha$
values ($> 45^{\circ}$) the current sheet part reaching the polar
cap is small; because of the limited extent of this region, whose
particle trajectories provide all the radiation, the
equatorial current sheet emission is far from uniform, something
that affects the radiation pattern on the sky-map.

\subsection{Comparison with Fermi Observations}

In Figs.~\ref{fig08}, \ref{fig09} and \ref{fig10} we plot
light-curve atlases for $\sigma=0.02\Omega, 30\Omega$, and
$1000\Omega$, respectively. Each atlas presents the light curves for
the indicated inclination and observer angle ($\alpha, \zeta$)
values. For $\sigma=0.02\Omega$ (Fig.~\ref{fig08}) we see single and
double peaked light-curves but the pulses seem to be rather wide. In
general, the light curves become double peaked as we go towards high
$\alpha$ and/or $\zeta$ values. For $\sigma=30\Omega$
(Fig.~\ref{fig09}) there are cases (mostly near high $\zeta$ values)
where the light-curves are quite narrow. However, there are cases
where the light-curves are still wide and in some cases they look
more complicated. This happens because for this level of $\sigma$
values there are different emitting components from both the inner
(within the LC) and the outer (outside the LC) magnetosphere. When
an observer is in a direction ($\zeta$ value) that sees both
components the corresponding light-curve becomes more complex. For
much higher $\sigma$ all the emission comes from the outer
magnetosphere outside the LC in regions near the equatorial current
sheet. Figure~\ref{fig10} shows that most of the light-curves, for
$\sigma=1000\Omega$, are narrow. Moreover, in this case the
observers of relatively small $\zeta$ values do not cross any
significantly emitting region and so the corresponding part of the
atlas is empty\footnote{The smaller the $\alpha$ value is, the
higher the $\zeta$ value below which the signal is absent.}.
{`Significantly emitting' in this case means that the maximum
luminosity per unit solid angle for the specific $\zeta$ is larger
than 2\% of the corresponding maximum value of the entire sky-map.
Below that value we don't get clear and reliable information. In
some of these cases it is also difficult to determine a clear
primary and/or a secondary local maximum.} Above that $\sigma$ value
the light-curve shape saturates and remains almost the same.

Some of the light-curve shapes shown in
Figs.~\ref{fig08}-\ref{fig10} resemble {those observed while others
do not.} However, in the vast majority of the observed pulsars we
have no clue about either their inclination angle ($\alpha$) or our
observer angle ($\zeta$) and so it is not easy to say whether a
specific pulsar is described and up to what level by each model.
Thus, we need a statistical comparison between the results provided
by our models and the observational data. In this paper we are
interested in checking the light-curve phenomenology, leaving for a
forthcoming paper the study of the energetic part of the high energy
emission.

\begin{figure*}
  \centering
  \includegraphics[width=\textwidth]{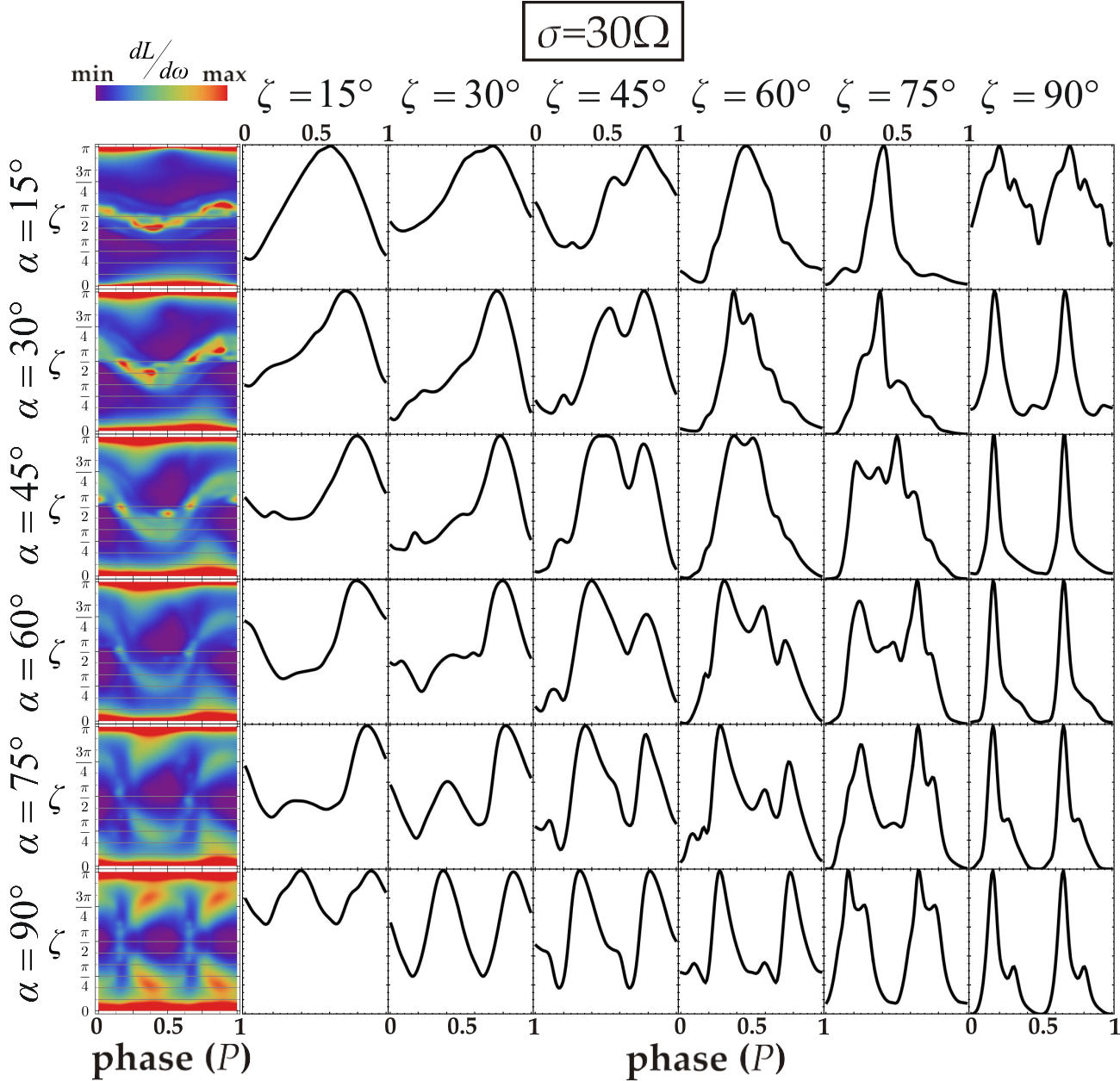}\\
  \caption{Same as in Fig.~\ref{fig08} but for $\sigma=30\Omega$. A significant part
  of the emission is produced in regions near the equatorial current sheet that
  has formed outside the light-cylinder. Some light-curves appear
  complex since in these cases the corresponding observer sees emission
  coming from different parts (inner and outer) of the magnetosphere.}\label{fig09}
\end{figure*}

\begin{figure*}
  \centering
  \includegraphics[width=\textwidth]{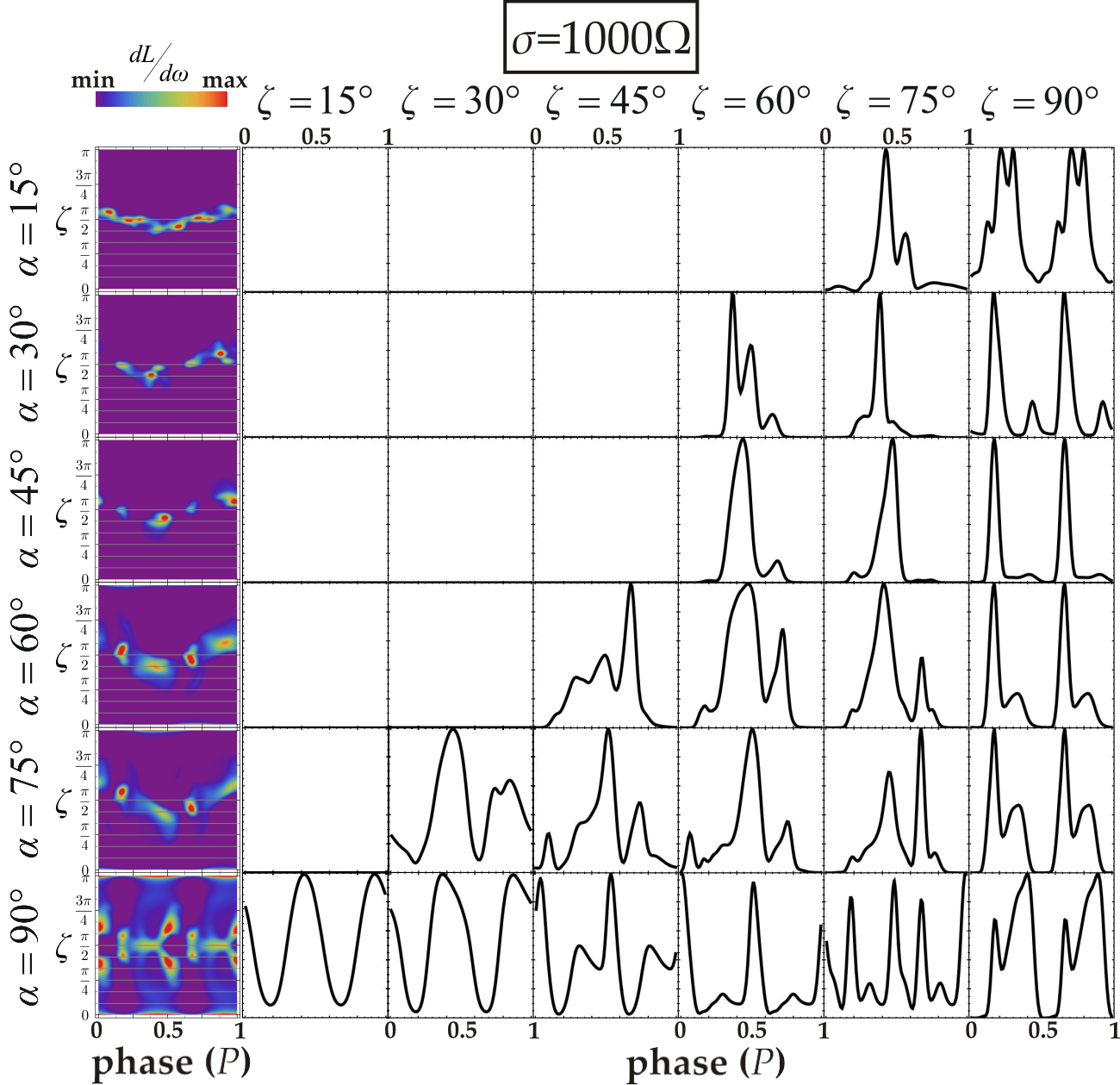}\\
  \caption{Same as in Figs.~\ref{fig08} and \ref{fig09} but for $\sigma=1000\Omega$.
  All the emission is produced in regions near the equatorial current sheet
  outside the light-cylinder. However, the emissivity is not uniform in these
  regions due to the modulation of the parallel electric components remaining within
  the light-cylinder. Nonetheless, the light-curves are, in general, narrow.
 The empty panels in this atlas are for $\zeta$ values where
  no radiation is observed for the corresponding $\alpha$ values.}\label{fig10}
\end{figure*}

As we mention in the first section, a robust statistical result is
the correlation of the $\gamma$-ray peak-separation $\Delta$ and the
radio-lag $\delta$ between the $\gamma$-ray and radio emission.
Figure~\ref{fig11} shows the $\delta$ vs $\Delta$
plot\footnote{Phases $\delta,~\Delta$ are measured as a fraction
$[0,\ldots,1]$ of the pulsar period ($P$). Whenever, the
corresponding $\gamma$-ray light-curve has only one peak $\Delta$ is
set to 0.} of the Second {\em Fermi} Pulsar Catalog (2PC) data
\citep{2013ApJS..208...17A} together with the data of the
dissipative models of the indicated $\sigma$ values assuming that
the radio emission comes from a region near the magnetic pole on the
pulsar surface. This assumption for the radio phase neglects phase
shifts (that would increase $\delta$) due to aberration and
retardation that result from emission at a significant altitude
relative to $R_{\rm LC}$. {In reality, the observed statistics in
$\delta-\Delta$ diagram is expected to depend on the details of both
the radio and $\gamma$-ray emitting models and on the possible
population biases of the radio and $\gamma$-ray pulsars. Thus, for
instance, the probability for radio photons to be observed is
expected to be higher (lower) for low (high) $|\zeta-\alpha|$
values. However, in the current study we don't incorporate any model
of radio emission. Thus, all the radio-lag $\delta$ values presented
below measure the time delays between $\gamma$-ray photons and radio
photons which are thought to be emitted on the $(\boldsymbol{\mu},
\boldsymbol{\Omega})$ plane near the stellar surface (assuming that
the latter are always observed\footnote{This seems to be the case
for high spin-down pulsars.}).} We decided to plot and compare the
2PC data corresponding only to standard pulsars (and not the
millisecond pulsars) since these objects (standard pulsars) present
clearer and stronger indications that the radio emission is produced
near the stellar surface. The {2PC} data are plotted with {black
dots} with error-bars while the model data are plotted with various
shapes and colors. The color denotes the $\alpha$ value according to
the indicated color scale (blue, low $\alpha$ to red, high $\alpha$)
and the shape denotes the corresponding $\zeta$ value according to
the indicated shape scale (full circle, low $\zeta$ to horizontal
line, high $\zeta$). {The $\Delta$ value is calculated as the phase
difference (measured as a fraction of the stellar period, $P$)
between the two highest local maxima. We note that any local maximum
lower than $0.05$ of the value of the highest (first) maximum is not
counted as a local maximum and the corresponding light curve is
considered to have only one peak. The $\delta$ value is derived as
the phase of the first peak of the $\gamma$-ray peak (assuming $0$
to be the phase emitted at the magnetic pole on the stellar
surface). We note that the phase $\phi_{ph}$ of a photon emitted at
a point $A$ along the trajectory of a particle (integrated in the
inertial frame) is given by
\begin{equation}
\label{photonphase} \phi_{ph}=\left(\Omega
t_{A}-\phi_{\mathbf{v_{A}}}-\frac{\mathbf{r_{A}}\cdot\mathbf{v_A}}{v_{A}}\frac{1}{R_{\rm
LC}}\right)\negthickspace\negthickspace\negthickspace\mod 2\pi
\end{equation}
where $t_A$ is the integration time corresponding to the point $A$
(assuming that the integration starts at the stellar surface),
$\mathbf{v_A}, \mathbf{r_A}$ are the particle velocity and position
vectors at $A$, and $\phi_{\mathbf{v_A}}$ is the azimuth angle of
the velocity $\mathbf{v_A}$ with respect to the magnetic axis at
$t=0$ oriented according to $\mathbf{\Omega}$. The last term in
Eq.~\eqref{photonphase} formulates the light travel time delay.}

For low $\sigma$ values ($\sigma=0.02\Omega$ and $\sigma=1\Omega$)
we have run 6 models (every $15^{\circ}$,
$\alpha=15^{\circ},~30^{\circ},\ldots,90^{\circ}$) and we have
calculated the $(\delta,~\Delta)$ values for 6 $\zeta$ values
$(\zeta=15^{\circ},~30^{\circ},\ldots,90^{\circ})$ while for
$\sigma>1\Omega$ we have run 17 models (every $5^{\circ}$,
$\alpha=10^{\circ},~15^{\circ},20^{\circ},\ldots,90^{\circ}$) and we
have calculated the $(\delta,~\Delta)$ values for 17 $\zeta$ values
($\zeta=10^{\circ},~15^{\circ},20^{\circ},\ldots,90^{\circ}$). This
means that the plots of Fig.~\ref{fig11} for $\sigma\leq1\Omega$ may
have up to 36 model data points while those for $\sigma>1\Omega$ may
have up to 289 model data points.

\begin{figure*}
  \centering
  \includegraphics[width=\textwidth]{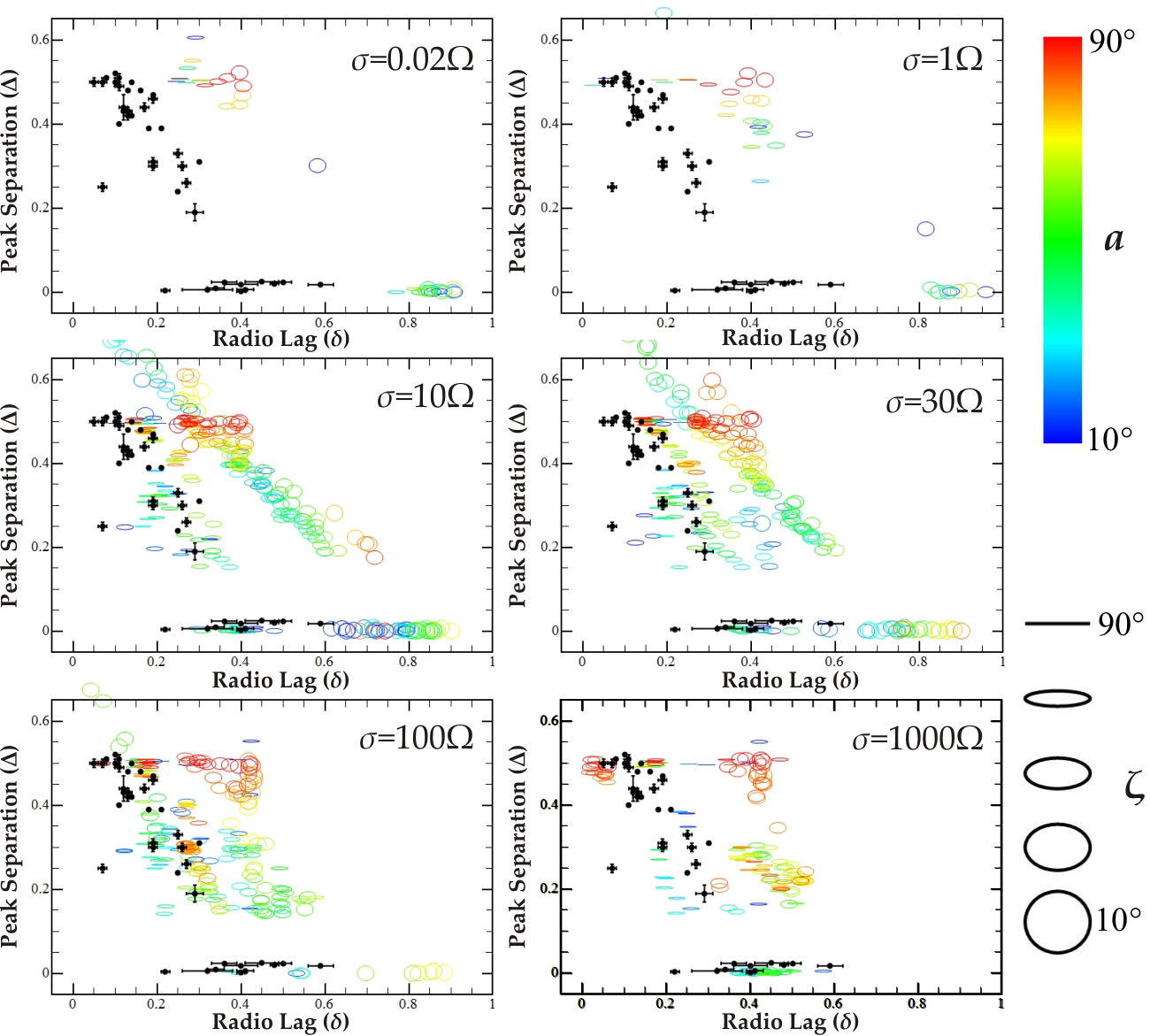}\\
  \caption{Diagram showing radio-lag $\delta$ vs the $\gamma$-ray
  peak-separation $\Delta$ for the models (color elliptical points)
  corresponding to the indicated $\sigma$ values. In each panel,
  the black points (with error-bars) correspond to {standard} pulsars
  (and not the millisecond ones) observed by {\em Fermi} (2PC).
  The color and the shape of the model points
  denote the $\alpha$ and $\zeta$ values, respectively as indicated in the figure.
  We note that for $\sigma=0.02\Omega$ and $\sigma=1\Omega$ we have considered
  a smaller number of $\alpha$ and $\zeta$ values than those for the higher
  $\sigma$ values (see text for more details). The low $\sigma$
  models have the highest radio-lag values among all the models. In this
  case all the model points lie in regions absent of observed points. As $\sigma$
  increases the model points move toward smaller radio-lag values
  (closer to the observed values). For very high $\sigma$ values many model points
  lie near the observed ones while there are still points, corresponding mostly to
  high $\alpha$ values, that lie in regions not covered by the observed ones.}
  \label{fig11}
\end{figure*}

The values of ($\delta,~\Delta$) can be determined from the shapes
of the computed model pulses. However, because this procedure is
time consuming and possibly biased, we have developed an algorithm
that takes into account most of the human-eye criteria employed in
the determination of ($\delta,~\Delta$) values. This allowed an
{automatic, unbiased and fast} computation of the ($\delta,~\Delta$)
values directly from the corresponding sky-map data. Such an
automatic calculation may lead to some errors (false points).
However, we expect the number of these false points to be small
since we checked more than a hundred light-curves of various types
and in more than 90\% of the cases the results were in total
agreement with those derived by simple eye examination. These false
points come from non-standard light-curves (more than two clear
peaks from more than one radiating components) and do not affect the
underlined statistics.

\begin{figure*}
  \centering
  \includegraphics[width=\textwidth]{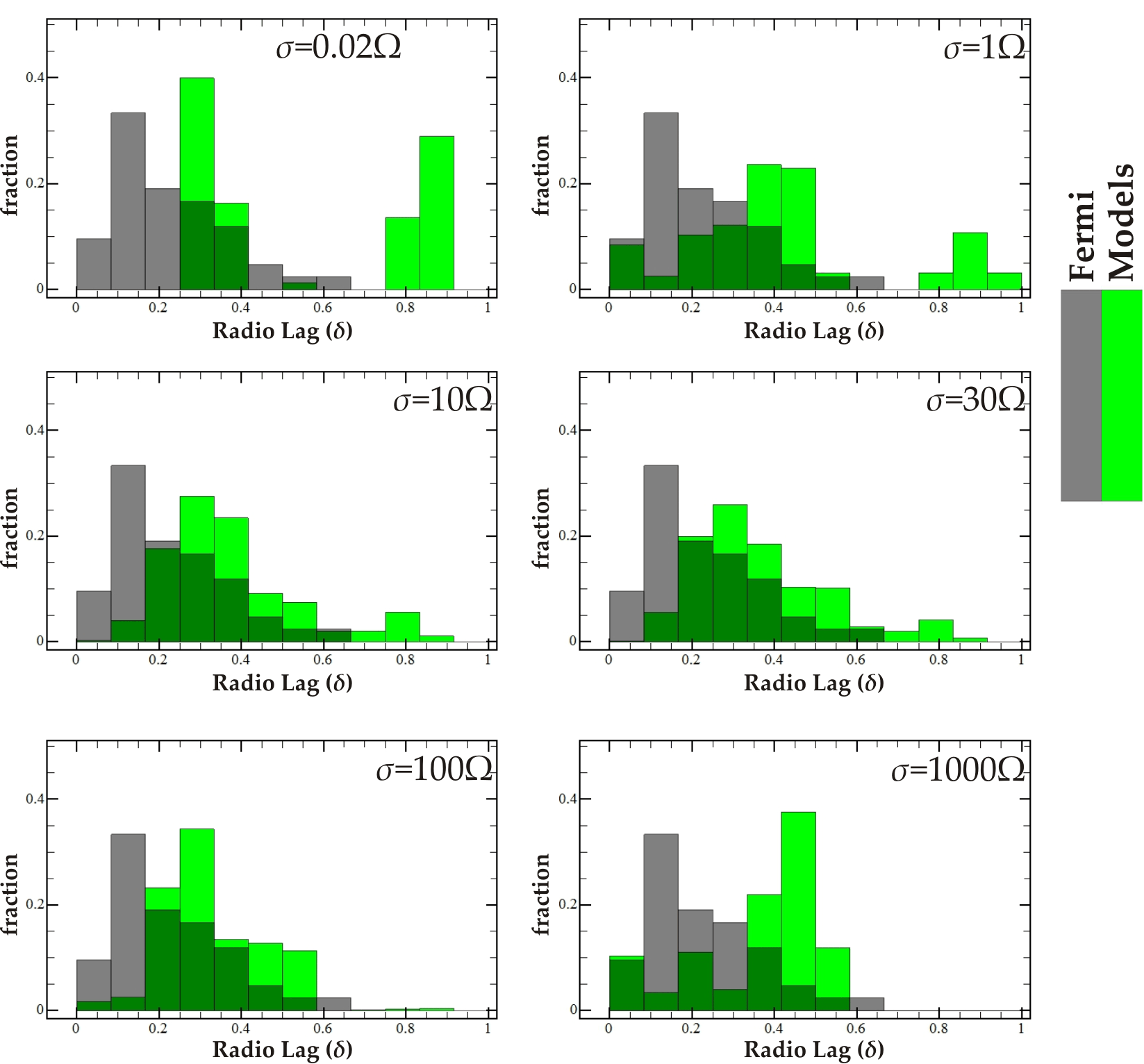}\\
  \caption{Histograms of the fraction of observed points (gray color)
  and model points (light green color) at a given radio-lag $\delta$, for the indicated $\sigma$
  values, along the radio-lag $(\delta)$ axis of Fig.~\ref{fig11}. The
  dark green color indicates coexistence of two histograms.
  The histograms for the models depend on the results presented in
  Fig.~\ref{fig11}. However, the results of Fig.~\ref{fig11}
  correspond to uniform distribution functions for $\alpha$ and
  $\zeta$ while these histograms have been calculated assuming
  a uniform distribution function for $\alpha$ and spherically uniform
  distribution function for $\zeta$. We see that low $\sigma$ models
  do not have good agreement with the observations. Higher $\sigma$ models cover the entire range
  of the observed $\delta$ values but the distributions differ from the
  observed ones.}\label{fig12}
\end{figure*}

In Figs.~\ref{fig12} and \ref{fig13} we plot the projected fraction
distributions (in green histograms) of the models shown in
Fig.~\ref{fig11} along the $\delta$ and $\Delta$ axes, respectively,
keeping the axis orientation of Fig.~\ref{fig11}. Each panel shows
also the fraction distributions of the observed points
(in gray histograms). We note that the model statistics shown in
Figs.~\ref{fig11}-\ref{fig13} depend not only on the {$\sigma$
value} but also on the {probability distribution functions of
$\alpha,~(F_{\alpha})$ and $\zeta,~(F_{\zeta})$}. Apparently, the
allocation of model points in Fig.~\ref{fig11} corresponds to
uniform distributions for $\alpha$ and $\zeta$. For $\alpha$ the
intrinsic {probability distribution function} $F_{\alpha}$ is
unknown but for $F_{\zeta}$ there is no reason why this would be
different from a spherically uniform $F_{\zeta}\propto \sin(\zeta)$.
Thus, the histograms in Figs.~\ref{fig12},~\ref{fig13} are modified
with respect to the data of Fig.~\ref{fig11} so that they are in
compliance with a uniform distribution in
$\alpha,~F_{\alpha}(\alpha)\propto 2/\pi$ and spherically uniform
distribution in $\zeta,~F_{\zeta}(\zeta)\propto \sin(\zeta)$.

\begin{figure*}
  \centering
  \includegraphics[width=\textwidth]{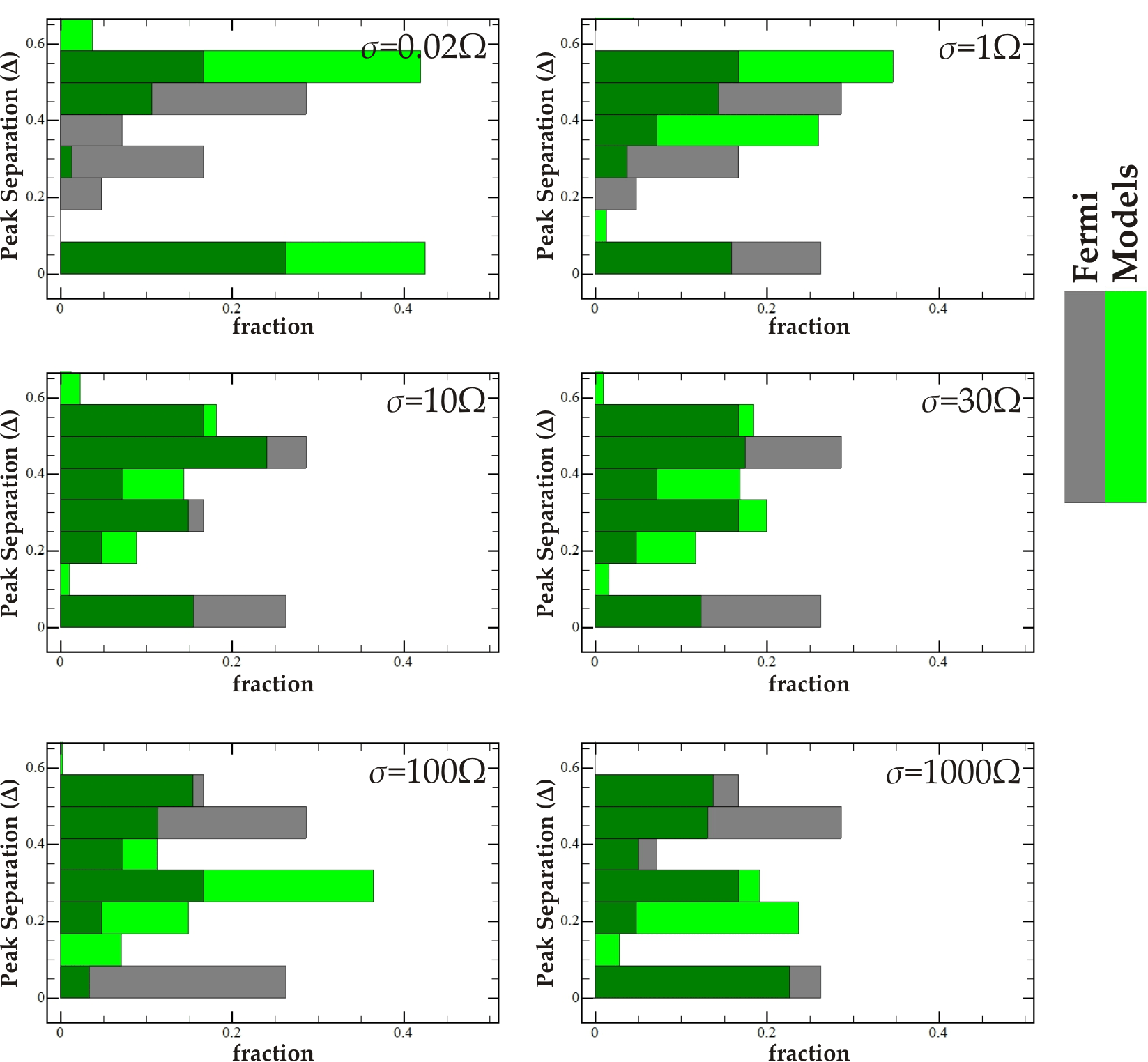}\\
  \caption{Same as in Fig.~\ref{fig12} but for $\Delta$ axis. Note that
  we have kept the axis orientation shown in Fig.~\ref{fig11}.}\label{fig13}
\end{figure*}

Figures~\ref{fig11}-\ref{fig13} show that for $\sigma=0.02\Omega$
there are mostly two groups of points (one near $\Delta=0$ and one
near $\Delta=0.5$) that have much higher $\delta$ values than the
observed ones. The group near $\Delta=0.5$ consists in general of
points coming from higher $\alpha$ values than those we see in the
group near $\Delta=0$. For $\sigma=1\Omega$ we still see the same
two groups of points but now some of the points of the group near
$\Delta=0.5$ have moved either towards smaller $\delta$ values or
smaller $\Delta$ values trying to fill the gap between the two
groups. For a higher $\sigma$ ($\sigma=10\Omega$) many model points
have moved to even smaller $\delta$ and $\Delta$ values overlapping
the region covered by the observed points while many model points
lie along a diagonal line, parallel to the arrangement of the
observed points, at higher $\delta$ values. As $\sigma$ increases,
more model points lie in the observed region while the points
remaining along the diagonal approach slightly the observed region
($\sigma=30\Omega$) and become more dispersed ($\sigma=100\Omega$).
Above this $\sigma$ value the total number of model points decreases
since for some $\zeta$ values the pulsars become invisible (see
Fig.~\ref{fig10}). This effect affects the histograms by increasing
the relative fractions at higher $\delta$ and lower $\Delta$.
Nonetheless, for higher $\sigma$ the model point distributions
saturate consisting always of points that lie in the observed region
and points that clearly lie in regions absent of observed points. We
note that the latter model points correspond mainly to higher
$\alpha$ values.

\begin{figure*}
  \centering
  \includegraphics[width=\textwidth]{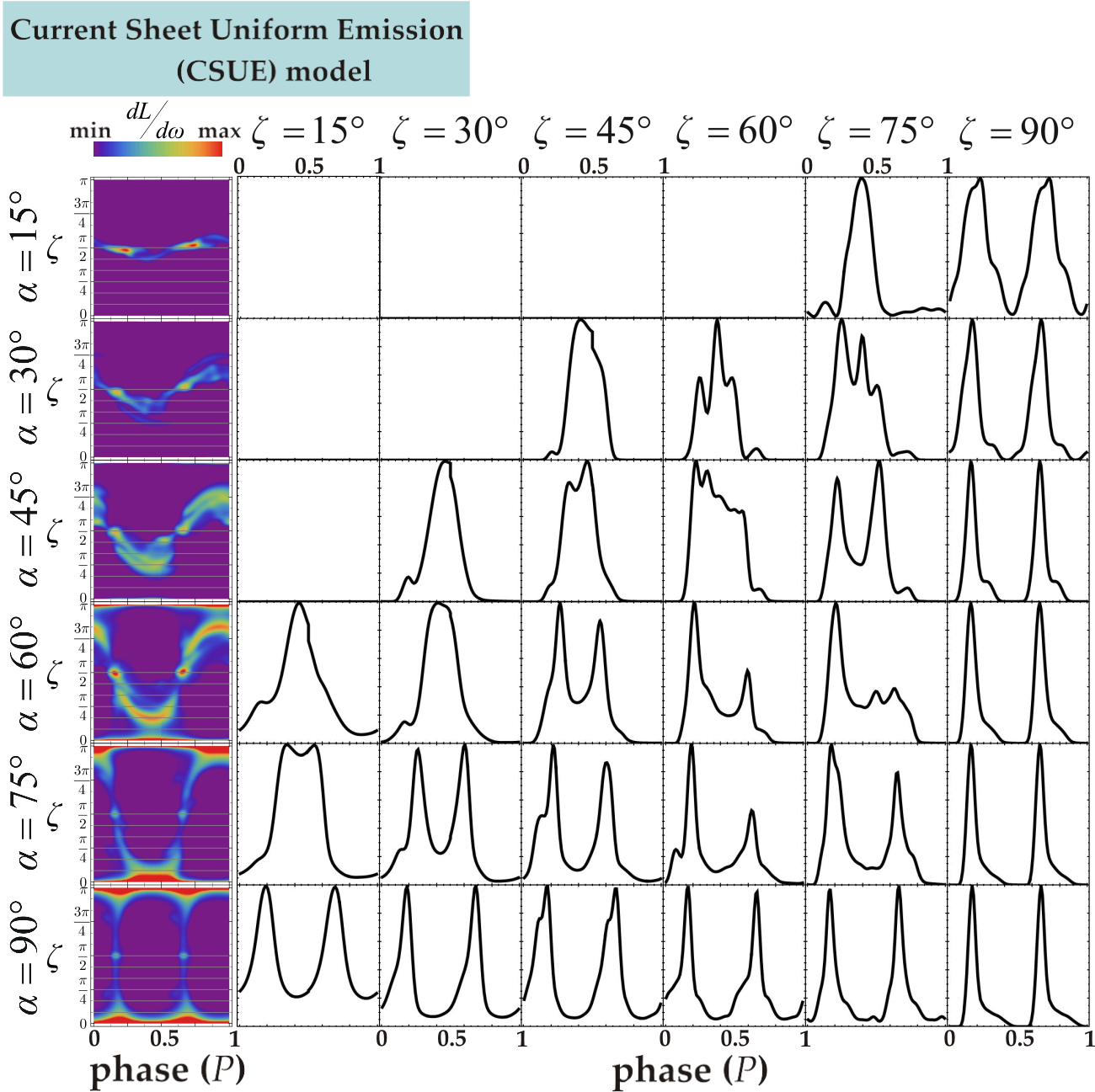}\\
  \caption{{Same as Fig.~\ref{fig14} but for the CSUE models.
  We see a variety of light-curve shapes similar to those
  observed by \emph{Fermi}.}}\label{fig135}
\end{figure*}

\section{Successful Models}

The results presented in the previous section, provide the basic
properties and trends of the radiation patterns in dissipative
models of uniform $\sigma$ in the open field line region. This study
showed that the low $\sigma$ models are consistent with inner
magnetosphere emission (within the LC) while the high $\sigma$
models (near-FFE) are consistent with outer magnetosphere emission
in regions near the equatorial current sheet outside the LC.
Moreover, the detailed comparison with the observed phenomenology
seems to rule out (under the assumption that the radio emission is
produced near the magnetic poles) the very low $\sigma$ models since
these models appear {to have} systematically larger radio-lags than
the observed ones contrary to what the application of the
theoretical accelerating models \citep[Outer-Gap, hereafter, OG;
Slot-Gap, hereafter, SG;][]{chr1986,romyad1995,mushar2004} on
dissipative magnetosphere solutions has shown
\citep{bs10b,2011arXiv1111.0828H,2012ApJ...754L...1K}. In fact, the
theoretical accelerating models under the geometry of near-vacuum
solutions indicate smaller radio-lags than those we get under the
geometry of near-FFE solutions. {However, the theoretical
acceleration models in this case (near-vacuum solutions) assume the
presence of significant $E_{\parallel}$ in regions where they do not
actually exist.} On the other hand, the high $\sigma$ models
(near-FFE) signify the importance of the regions near the equatorial
current sheet and are clearly closer to observations. {They show a
set of light curves that fit the observations while another set has
larger radio-lags than those observed.} This inconsistency leaves
open questions on whether the {uniform $\sigma$} models are able to
reproduce the observed $\gamma$-ray light-curve phenomenology.
Nevertheless, as we will see, the study above reveals a way to find
the successful models.

Up to now we have assumed a uniform {$\sigma$ distribution in the
open field regions (the closed field line region has always high
$\sigma$)} and that the radio emission is produced near the magnetic
poles. The uniform $\sigma$ is the simplest and {most unbiased
approach since it gives a uniform level of screening of
$E_{\parallel}$ everywhere.} This means, especially for the high
$\sigma$ (near-FFE) solutions, that the screening of the
$E_{\parallel}$ is more difficult wherever the required FFE
$(\boldsymbol{\nabla}\times\mathbf{B})_{\parallel_{\text{FFE}}}$ is
higher. We note that this is, in general, valid for all the current
density $\mathbf{J}$ prescriptions tested in this study.

\begin{figure*}
  \centering
  \includegraphics[width=12cm]{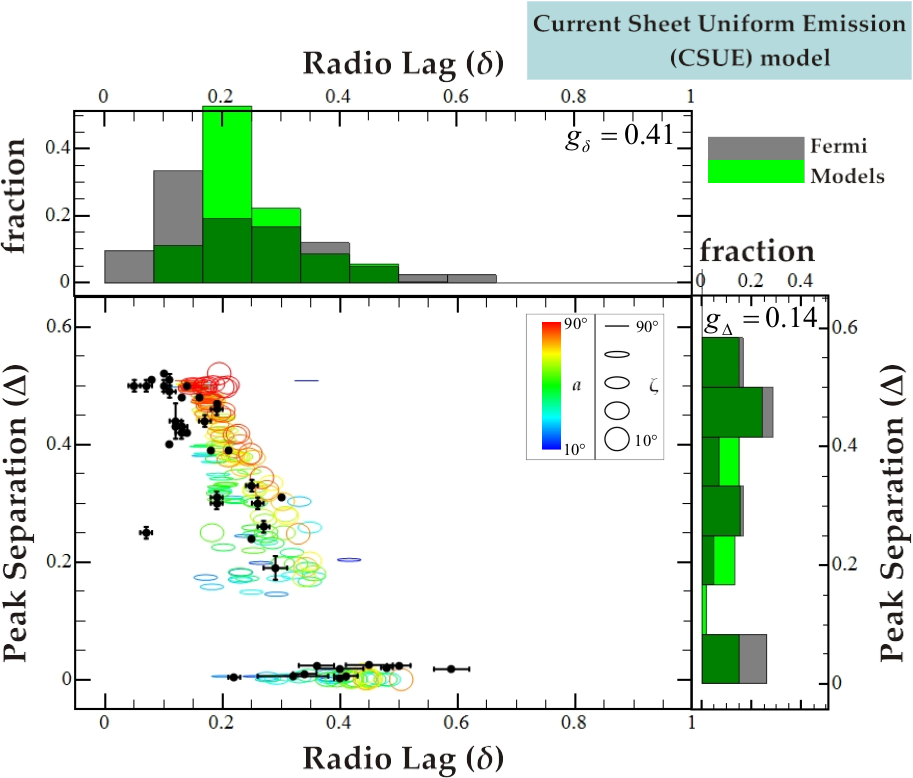}\\
  \caption{The $\delta$-$\Delta$ diagram for the Current Sheet
  Uniform Emission (CSUE) model together with the corresponding histograms
  along $\delta$ (top) and $\Delta$ (right) axes. Even though these models
  are in a better agreement with observations than the dissipative models
  shown in Figs.~\ref{fig11}-\ref{fig13} they still present a systematic
  trend of model radio-lag $(\delta)$ values that are higher than those observed.
  In each histogram panel we have indicated the corresponding $g$ values
  derived by relation \eqref{minimeq} {for uniform $\alpha$ distribution
  (all $w_{\alpha_i}$ are equal).}}\label{fig14}
\end{figure*}

\begin{figure}
  \centering
  \includegraphics[width=5cm]{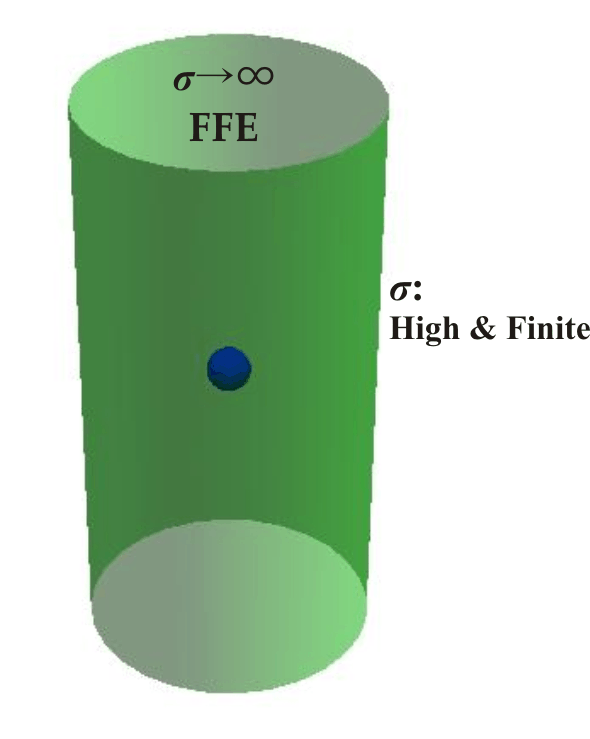}\\
  \caption{A simplified picture that describes the structure of the
  models successfully reproducing the observed $\gamma$-ray
  light-curve phenomenology. The blue sphere and the green
  cylindrical surface represents the pulsar and the light-cylinder,
  respectively. This type of model assumes a FFE regime Inside the
  light-cylinder and Dissipative Outside (FIDO).
  }\label{fig15}
\end{figure}

{Assuming that the flow across the polar cap is not highly
non-uniform} the discrepancy between the model data and the {\em
Fermi} data, present even for the highest $\sigma$ values shown in
Figs.~\ref{fig11}-\ref{fig13}, could indicate primarily two things:
\newline Either
\begin{enumerate}[(a)]
    \item that the radio emission is not produced near the magnetic
    poles

\end{enumerate}
or
\begin{enumerate}[(b)]
    \setcounter{enumi}{1}
    \item that the adopted $\sigma$ distribution {and/or
        the adopted Ohm's law are not the correct ones.}
\end{enumerate}

Case (a) is in contradiction with a widely observed phenomenology
\citep[e.g.][]{1983ApJ...274..333R} and we will not consider this
option in this paper\footnote{{We note, however, that there are
cases (e.g. Crab pulsar, some millisecond pulsars) where the radio
emission is in phase with the $\gamma$-ray emission. This fact
indicates for these cases the possibility that the radio emission is
also produced in the outer magnetosphere where the $\gamma$-rays are
produced.}}. The second {case (b)} seems reasonable and it should be
related to the detailed properties of the underlined microphysics
{of pair creation}. Nonetheless, the analysis presented in
Section~\ref{sec_results} indicates that the non-uniform
``lighting'' of different parts of regions near the equatorial
current sheet may produce different radiation patterns on the
sky-maps. \cite{ck2010}, \cite{bs10b}, \cite{2011MNRAS.412.1870P}
and \cite{2013A&A...550A.101A} have already presented $\gamma$-ray
light-curves and studied some of their phenomenological properties
assuming either constant emission or constant $\gamma_L$ values on
the equatorial current sheet. However, only \cite{ck2010} attempted
a coarse comparison with the observed {phenomenology in the First
{\em Fermi} Pulsar Catalog (1PC)}.

\begin{figure*}
  \centering
  \includegraphics[width=12cm]{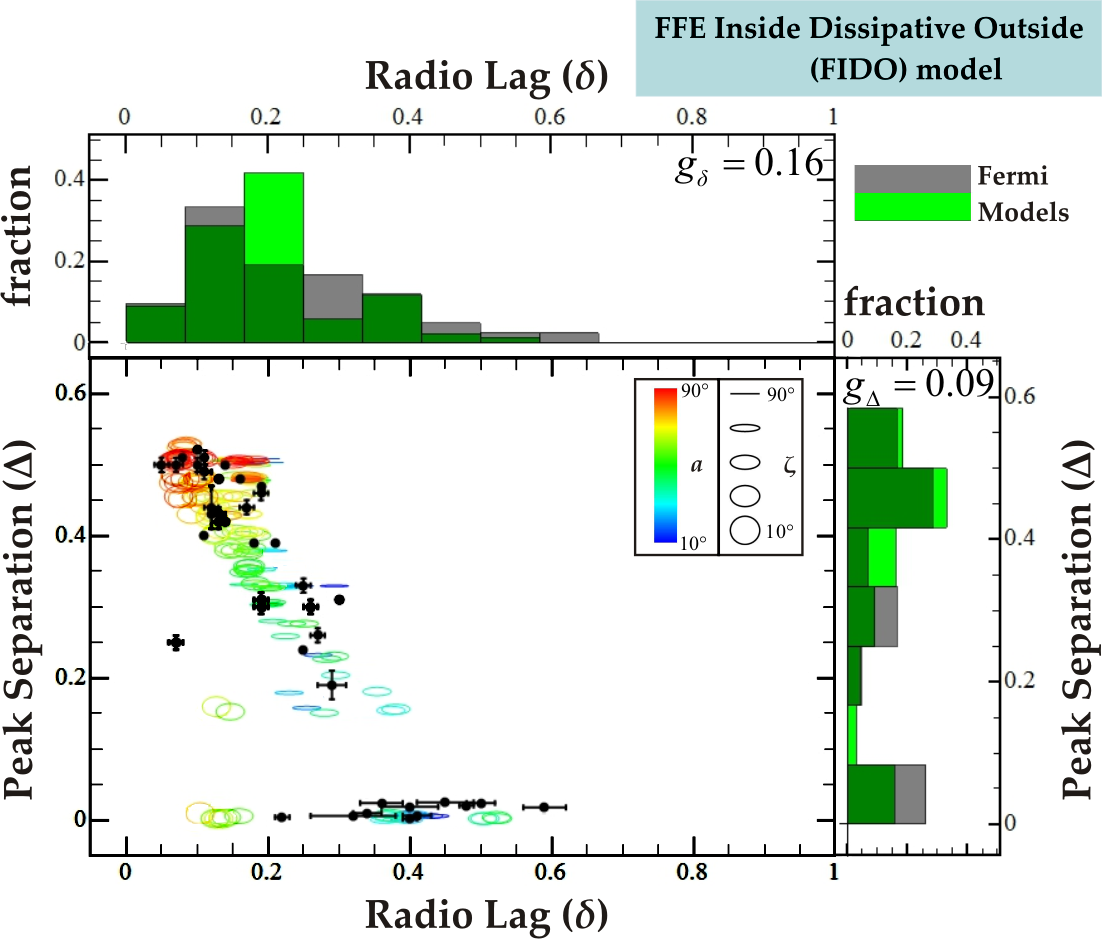}\\
  \caption{Same as Fig.~\ref{fig14} but for the FIDO models. We see that
  these models are in a very good agreement with the observations.
  The indicated $g$ values are considerably smaller than those of CUSE models
  (Fig.~\ref{fig14}).}\label{fig16}
\end{figure*}

We first make, for comparison, a detailed statistical analysis of
the $\delta-\Delta$ relation for the dissipative models, assuming
uniform emissivity everywhere on the equatorial current sheet of the
FFE solutions. For this we assumed that all the particles start from
the outer parts of both polar caps. More specifically, using the
Open Volume radial, $r_{\rm ovc}$, and azimuthal, $\phi_{\rm ovc}$,
Coordinates (OVCs) \citep{dykhar2004} defined inside the open volume
of each FFE solution, we assume that particles originate in a
uniform distribution within the thin layer between $r_{\rm ovc} =
0.95 - 1$ of the corresponding polar cap radius. We integrate the
trajectories of these particles up to $2.5R_{\rm LC}$ assuming
constant emissivity along their length. {Figure~\ref{fig135} shows
the sky-maps and the corresponding light curve atlas. These results
are very similar to the Separatrix Layer (SL) model in \cite{bs10b}.
Actually, our sky-maps seem a little more complex than those of
\cite{bs10b} mostly because we have considered uniform particle flux
over the entire part of the polar cap with $r_{\rm ovc} = 0.95 - 1$
instead of a flux from a narrow zone around $r_{\rm ovc} =0.90$ or
$0.95$.} In Fig.~\ref{fig14} we present the ($\delta, \Delta$)
statistics together with the corresponding histograms along $\delta$
and $\Delta$ axes, for the FFE (or near-FFE) models assuming uniform
emission (up to $2.5R_{\text{LC}}$) on the equatorial current sheet.
All these panels include {the 2PC} data for comparison.
Figure~\ref{fig14} shows that the Current Sheet Uniform Emission
model (hereafter, CSUE) is in better agreement with the observations
than the series of dissipative models presented in
Figs.~\ref{fig11}-\ref{fig13}. Nonetheless, it is apparent that
there is a strong trend toward higher radio-lag values $(\delta)$
compared to those observed. This difference becomes clearer in the
distribution along the $\delta$ axis.

\begin{figure*}
  \centering
  \includegraphics[width=\textwidth]{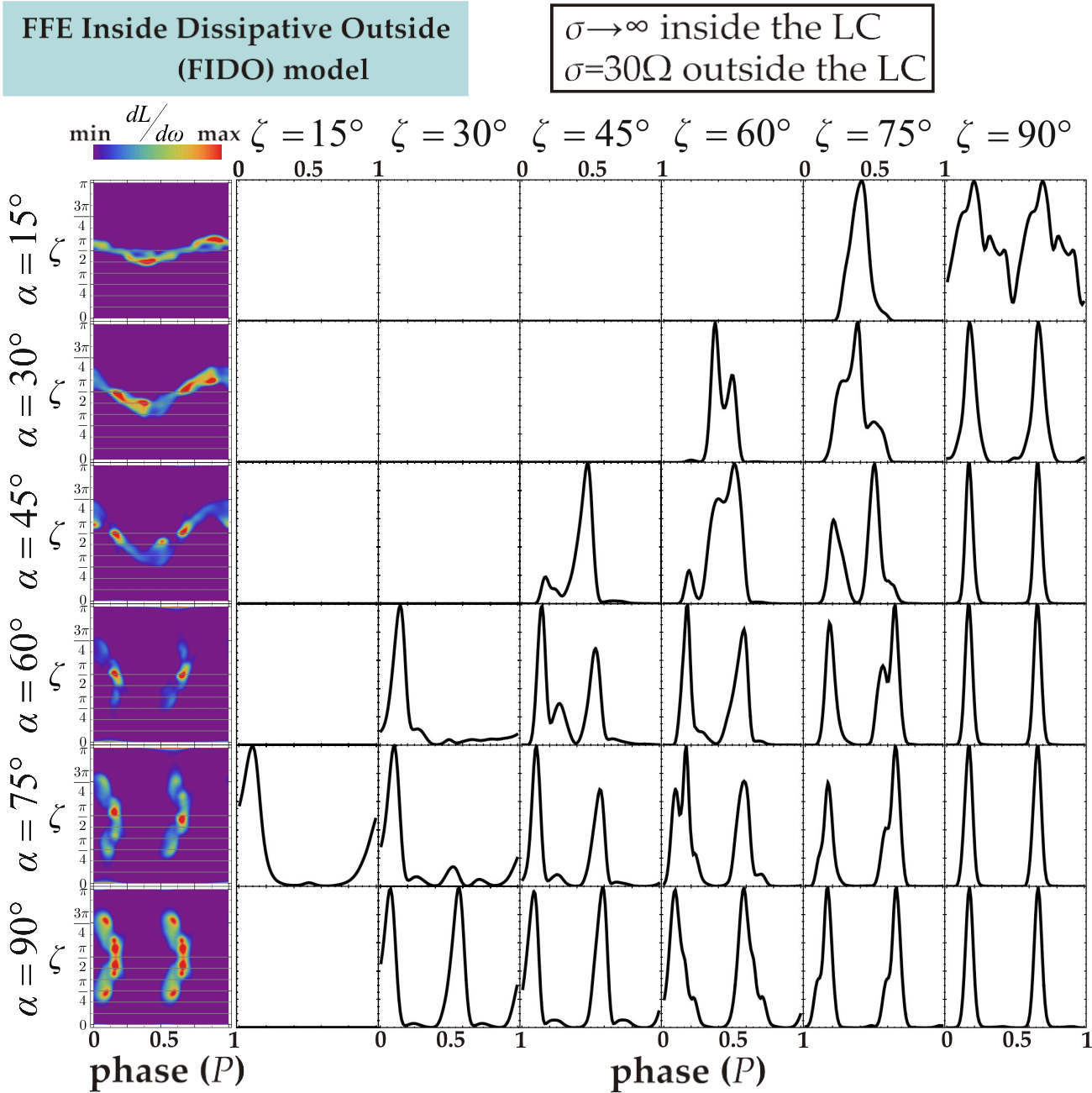}\\
  \caption{Same as Fig.~\ref{fig08} but for the FIDO models. The adopted
  conductivity value outside the light-cylinder is $30\Omega$. The emission
  comes from regions near the equatorial current sheet. The emissivity
  distribution is affected by the local physical properties and
  is not symmetric around the equatorial current sheet. This
  adjusts the corresponding radio-lag values toward the observed
  values. Moreover, there is no emission for low $\zeta$ and
  high $\alpha$ values. We see also narrow, well-defined $\gamma$-ray
  light-curves with low off-peak emission.}\label{fig18}
\end{figure*}

\begin{figure*}
  \centering
  \includegraphics[width=11cm]{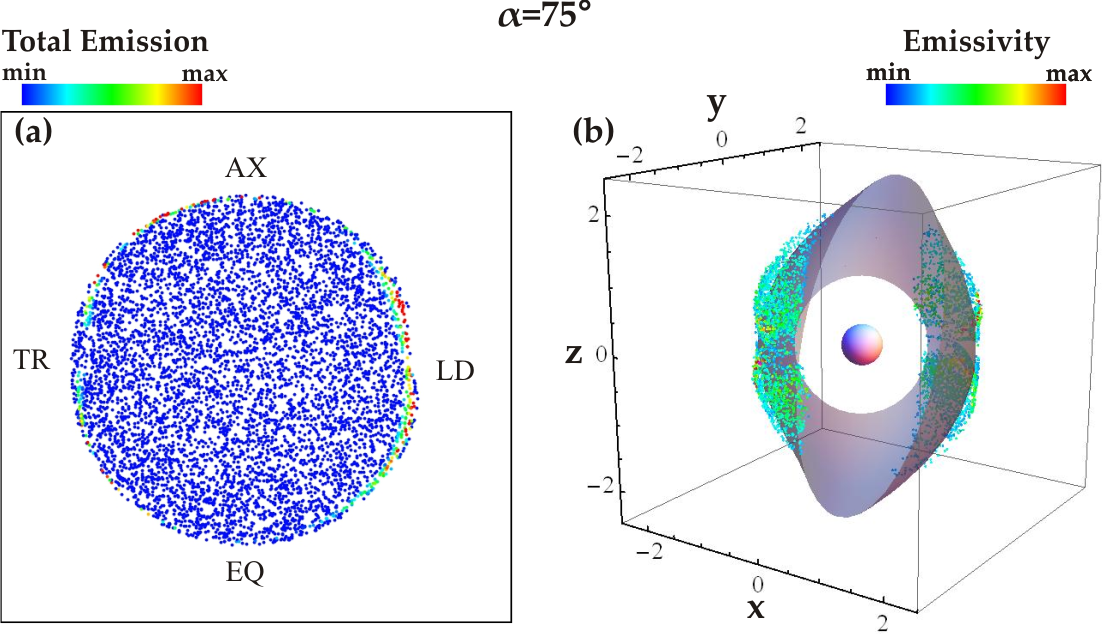}\\
  \caption{{\textbf{(a)} Same as the panels in the middle column of
  Fig.~\ref{fig07} but for the $\alpha=75^{\circ}$ FIDO model.
  We see that the high emission is produced mostly from particles
  that originate at the leading edge of the polar cap.}
  \textbf{(b)} Same as the panels of the right-hand column of Fig.~\ref{fig04}
  but for the $\alpha=75^{\circ}$ FIDO model. We see that most of the
  emission is produced in regions near the equatorial current sheet around
  the rotational equator.}\label{fig17}
\end{figure*}

On the one hand, the high $\sigma$ models presented in
Figs.~\ref{fig11}-\ref{fig13} fail because the emission from a
region near the equatorial current sheet is modulated by the
distribution of {$E_{\parallel}$ within the LC}. On the other hand,
when we assume constant emission, as in CSUE models, we still are
not able to reproduce the observed statistics. However, both cases
show that the emission from the outer magnetosphere emerging from
physical properties of the near-FFE solutions in regions near the
equatorial current sheet is important.

We consider an alternative simple model that retains the emission
near the equatorial current sheet and the physical properties of the
dissipative solutions but avoids its modulation from the inner
magnetosphere. We assume a hybrid FFE-Inside \& Dissipative-Outside
model (hereafter, FIDO) that has an FFE regime
$(\sigma\rightarrow\infty)$ within the LC and high (but finite)
$\sigma$ outside the LC (see Fig.~\ref{fig15}). {In this  model, the
inner magnetosphere no longer affects} the radiation patterns while
the finite $\sigma$ in the outer magnetosphere allows the regions
near the equatorial current sheet to determine the radiation
patterns based only on the local physical properties.
Figure~\ref{fig16} shows the $\delta-\Delta$ distribution
corresponding to these hybrid models for $\sigma=30\Omega$ outside
the LC together with the 2PC data\footnote{{We have considered the
FFE geometry with $E_{\parallel}$ that is provided by
Eq.~\ref{eparapprox} applying $\sigma=30\Omega$ for $R\geq R_{LC}$.
We have also checked that real dissipative solutions that have high
$\sigma$ $(>30\Omega)$ within the LC and $\sigma=30\Omega$ outside
the LC provide similar results (sky-maps, $\gamma_L$ values) when
any $E_{\parallel}$ these solutions sustain within the LC is
disregarded.}}. Remarkably, these models are by far the best in
reproducing the observed 2PC statistics. The high energy radiation
in this case is produced in regions near the equatorial current
sheet but the corresponding radiation patterns depend exclusively on
the local physical properties with no contribution from inside the
LC. This affects the radiation patterns of the models, with those
having the higher inclination angles $(\alpha> 45^{\circ})$ being
the most affected.

{Figure~\ref{fig18} presents the light-curve atlas of FIDO models.
We see narrow and well defined light-curves that resemble those
observed by {\em Fermi}. For low $\alpha$ values
$(\alpha\leq45^{\circ})$ the emission comes from regions near the
entire equatorial current sheet while for higher $\alpha$ values
$(\alpha>45^{\circ})$, where the equatorial current sheet extends to
lower colatitudes, the emission comes from regions near the
equatorial current sheet that are closer to the rotational equator.
This effect eliminates the corresponding emission at low $\zeta$
values. In Fig.~\ref{fig17}a we map, in the indicated color scale,
the total luminosity on the polar cap for $\alpha=75^{\circ}$
(similar to the middle column of Fig.~\ref{fig07}). We see that the
high emission trajectories lie near the edge of the polar cap. This
clearly indicates that all the emission is produced in regions near
the current sheet beyond the LC. However, the emission is not
uniform all along the polar cap edge. We observe stronger emission
from the leading edge of the polar cap and from parts that are
closer to the rotational axis. In Fig.~\ref{fig17}b we plot, in the
corotating frame, the points that trace the comoving (prescribed by
the particle flux) volume that contribute to the highest 95\% of the
total emission in the FIDO model of $\alpha=75^{\circ}$. We see that
most of the emission comes from a region near the equatorial current
sheet that has $\theta>45^{\circ}$ (where $\theta$ is the spherical
polar angle) towards the rotational equator. These regions have high
$E_{\parallel}$ and the particles that travel through them reach
higher $\gamma_L$ values.}

{The sky-map emission pattern depends, in general, on (a) the
distribution of the initial conditions ($r_{\rm ovc}, \phi_{\rm
ovc}$) of the highly emitting trajectories on the polar cap  and (b)
the altitude (distance) at which these trajectories emit.
\cite{bs10b} have already presented the projections of particle
trajectories on the sky-map based on the FFE geometry. Assuming
constant and uniform emission along all the trajectories they
concluded that the emitting patterns that match better the shapes of
the observed $\gamma$-ray light curves are those that originate
close to a ring corresponding to $r_{\rm ovc}=0.90$ or $r_{\rm
ovc}=0.95$ (for each $\phi_{\rm ovc}$ the edge of the polar cap has
$r_{\rm ovc}=1$). They noticed also that, in this case (FFE), the
formation of the caustics is the result of the Sky Map Stagnation
(SMS) effect which means that the strong emission in sky-maps is not
due to the coincidence that emission from different trajectories
congregates on the same point on the sky-map, but the emission from
different points of a single trajectory arrives simultaneously,
piling up on the same region of the sky-map. However, they noticed
(and we confirm) that this effect is less prominent the lower the
emission altitude is and the higher the $r_{\rm ovc}$ is.
Nevertheless, their results \citep[see fig.~6 and fig.~8
in][]{bs10b} indicate that high intensity regions on sky-maps can be
produced by the concurrent contribution from different trajectories
(especially for $r_{\rm ovc}\rightarrow 1$). In our case, instead of
setting the $r_{ovc}$ limits of the emitting trajectories
artificially, we allow the distribution of $E_{\parallel}$ from the
model to determine the emitting geometry. In reality, the emissivity
distribution among and along the trajectories is what will determine
the emission pattern on sky-maps. The geometric properties and the
effectiveness of the SMS phenomenon of the highly emitting
trajectories will modulate the sky-map emission patterns
accordingly.}

{In Fig.~\ref{fig185} we plot the sky-map projections of the
trajectories originating from one (north) polar cap for the
indicated $r_{\rm ovc}$ values (for the entire range of the
$\phi_{\rm ovc}$ values). The color along the trajectory projections
represents the distance $d$ traveled along each trajectory according
to the indicated color scale. These results confirm the general
picture presented by \cite{bs10b}. We see that the SMS effect
becomes stronger (i.e. shorter blue segments) as the $r_{\rm ovc}$
decreases. Figure~\ref{fig17}a and the bottom middle panel of
Fig.~\ref{fig07} show which trajectories presented in
Fig.~\ref{fig185} are the ones that shine brightest producing the
corresponding emission patterns on sky-maps.}

{The main differences that make FIDO models more effective in
reproducing the $\delta - \Delta$ correlation than CSUE models are:}
\begin{enumerate}[(a)]
    \item {In FIDO models the emission is non-uniform along
    the equatorial current sheet in a way that emission at small $\zeta$ is
    negligible because the regions, near the equatorial current
    sheet, producing radiation for small $\zeta$ have small
    $E_{\parallel}$ for high $\sigma$ and consequently the particles
    in these regions have low and inefficient $\gamma_L$ values. This effect
    removes some points of higher $\delta$ values from the $(\delta-\Delta)$
    diagram.}

    \item {In FIDO models the emission is asymmetric
        across the equatorial current sheet. The higher emission
        (i.e. higher $E_{\parallel}$) originates from the
        leading edge of the polar cap rim. These trajectories
        have smaller $\delta$ values that push some of the
        points of $(\delta-\Delta)$ diagram to left (smaller
        values of $\delta$).}
\end{enumerate}

\begin{figure*}
  \centering
  \includegraphics[width=\textwidth]{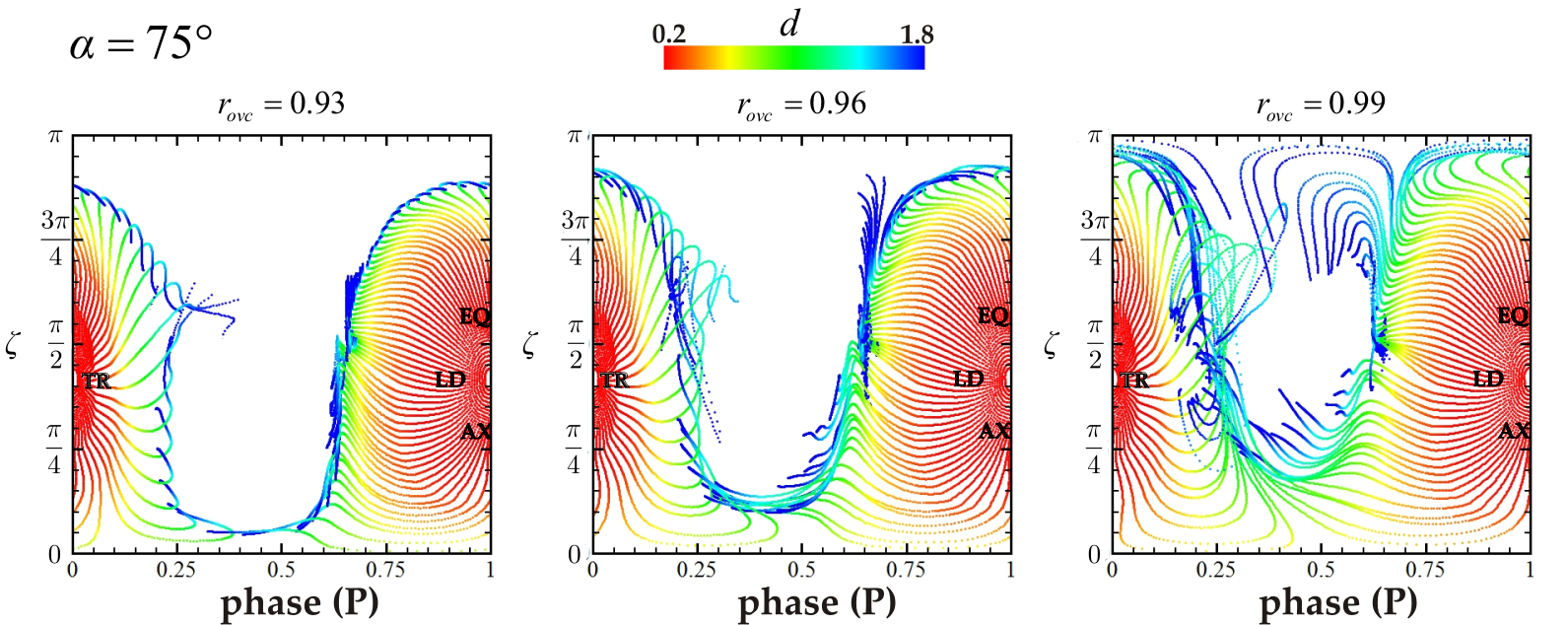}\\
  \caption{{Trajectory projections on the sky-map for $\alpha=75^{\circ}$.
  Each panel shows trajectories that originate from one (north) polar cap at the indicated
  $r_{ovc}$ values. The color along the trajectory projections illustrates the
  traveled distance $d$. Which trajectories emit and at what distance is what
  determines the emission pattern on the sky-maps.}}\label{fig185}
\end{figure*}

{The non-uniform and asymmetry effects were also present in the
dissipative models of uniform (in the open field line region)
$\sigma$ presented in Figs.~\ref{fig04}-\ref{fig13}. In the middle
column of Fig.~\ref{fig07} we see that the total emission near the
edge of the polar cap is not equal in diametrically opposed
points\footnote{The diametrical point of a point on the polar cap
rim.}. However, in that case these effects were modulated by the
$E_{\parallel}$ of the inner magnetosphere and produced different
emission patterns.}

{We have already mentioned that the statistical comparison between
the models (CSUE and FIDO) and the {2PC} observations presented in
Fig.~\ref{fig14} and Fig.~\ref{fig16} depends on the unknown
intrinsic probability distribution function of $\alpha$
$(F_{\alpha})$ and on the probability for the radio and $\gamma$-ray
pulses to be  observed together for the different $\alpha$ and
$\zeta$ combinations. However, the determination of this last
probability requires a detailed prescription for the radio emission
as well, something that goes beyond the scope of this paper.
Nonetheless, the determination of the $F_{\alpha}(\alpha)$ that
minimizes the statistical differences between the models and the 2PC
data on the $(\delta-\Delta)$ diagram is challenging. Thus, below we
do this exercise and determine the $F_{\alpha}(\alpha)$ functions
for the CSUE and FIDO models that fit best the data taking into
account all the points plotted in Figs.~\ref{fig14} and
\ref{fig16}.}

The distribution on the $(\delta-\Delta)$ plane is 2D and so any 2D
statistical comparison technique (e.g. Kolmogorov-Smirnov) is
unreliable due to the relatively small number of observational
points. Instead, we decided to find the relative weights for each
$\alpha$ value that minimize the differences between the
model-observation distributions, independently along the $\delta$
and $\Delta$ axis. In particular, we derived the values of
$w_{\alpha_i}: (\alpha_i=10^{\circ},~15^{\circ}\ldots 90^{\circ})$
that minimize the
\begin{equation}
\label{minimeq}
g_{\delta,~\Delta}(w_{\alpha_i})=\sum_{j=1}^{N_b}\frac{\left(\sum_{i=1}^{N_{\alpha}}w_{\alpha_i}
P_{\alpha_i}^j-P_O^j\right)^2}{\sum_{i=1}^{N_{\alpha}}w_{\alpha_i}
P_{\alpha_i}^j+P_O^j}
\end{equation}
under the requirements $w_{\alpha_i}\geq0: i=1\ldots N_{\alpha}$ and
$\sum_{i=1}^{N_{\alpha}}w_{\alpha_i}=1$. In Eq.~\eqref{minimeq}
$N_b$ is the number of bins we consider for the distributions along
the $\delta$ and $\Delta$ axis. The quantity $P_O^j$ expresses the
observable probability for the $j^{\rm th}$ bin while the quantity
$P_{\alpha_i}^j$ expresses the probability of models with
$\alpha=\alpha_i$ to be in the $j^{\rm th}$ bin. Finally $N_\alpha$
is the total number of different $\alpha$. The expression
\eqref{minimeq} is minimized {using {\em Interior Point} method as
this is implemented in \verb"Mathematica" \citep{mathematica}. We
note that the $g_{\delta}$ and $g_{\Delta}$ values are not compared
directly to each other since they refer to a different number of
bins $(N_b)$ along $\delta$ and $\Delta$ axis.

\begin{figure*}
  \centering
  \includegraphics[width=12.5cm]{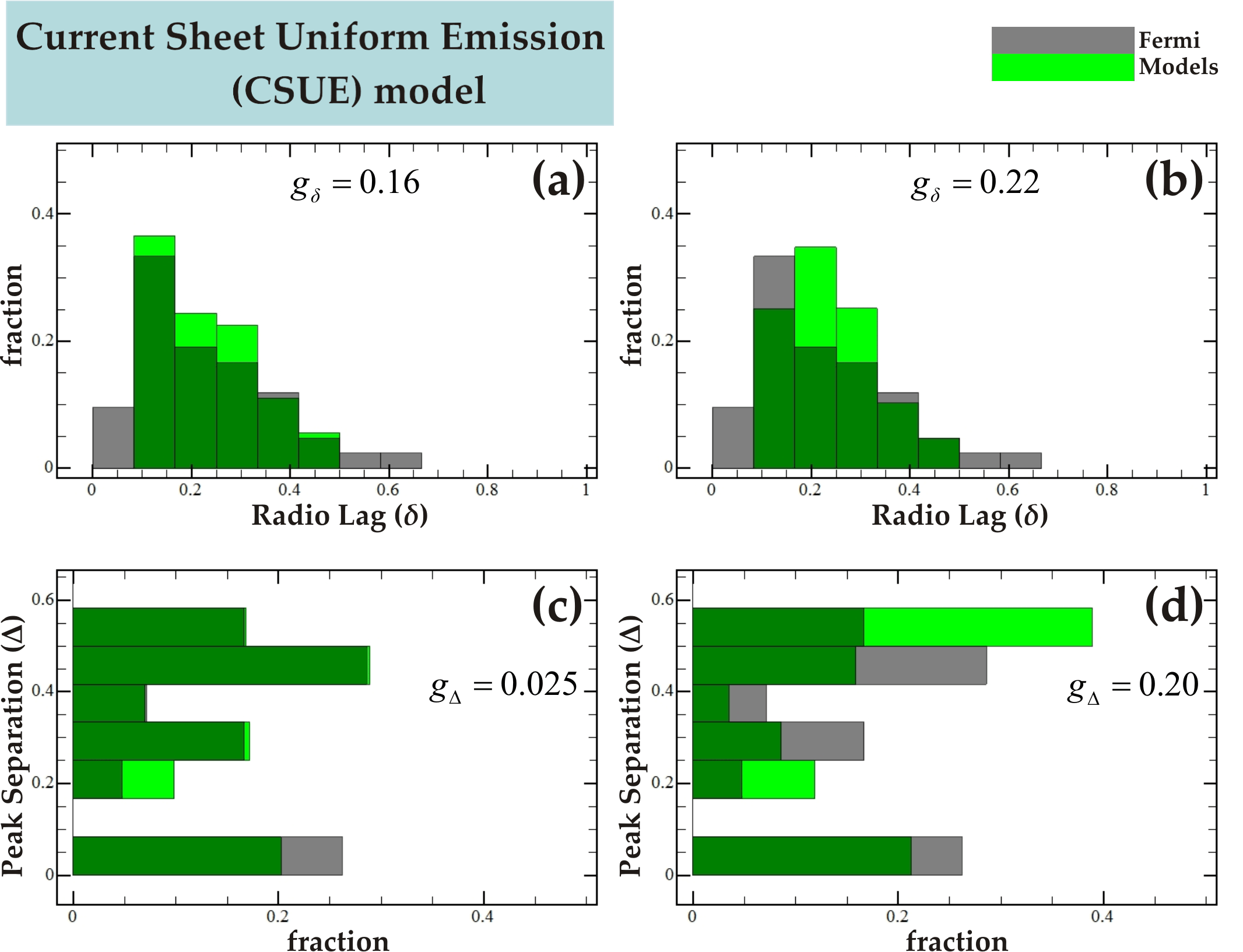}\\
  \caption{The CSUE model fraction histograms (together with the observed
  ones) for the optimum $\alpha$ distributions. \textbf{(a), (c)} The model
  histograms along $\delta$ and $\Delta$ axes for the (different) $\alpha$
  distributions that minimize the corresponding $g,~(g_{\delta},~g_{\Delta})$ values
  (Eq.~\ref{minimeq}), respectively. \textbf{(b), (d)} The model histograms
  along $\delta$ and $\Delta$ axes, respectively, for the $\alpha$
  distribution that is the mean of the $\alpha$ distributions that independently minimize the
  $(g_{\delta},~g_{\Delta})$ values in \textbf{(a), (c)}. In \textbf{(a), (c)}
  we see an improved performance for the CSUE models (with respect to Fig.~\ref{fig14})) even though they are still
  unable to reproduce the low radio-lag values.  When
  the corresponding optimum $\alpha$ distributions are combined \textbf{(b), (d)}
  there is poorer agreement (especially along the $\Delta$ axis),
 indicating that the optimum $\alpha$ distributions adopted in
  \textbf{(a)} and \textbf{(c)} are not very similar.}\label{fig19}
\end{figure*}

In Figs.~\ref{fig14},~\ref{fig16} we have indicated the $g$ values
corresponding to uniform $\alpha$ distributions (all $w_{\alpha_i}$
are equal). We see that FIDO models have $g$ values that are
{considerably} smaller than the (CSUE) models. This confirms our
previous conclusions that the FIDO models fit the observational data
quantitatively better than the CSUE models, under the assumption of
uniform $\alpha$ distribution.}

We calculated the $w_{\alpha_i}$ that minimize the $g(w_{\alpha_i})$
corresponding independently to the distributions along $\delta$ and
$\Delta$ axes for both CSUE and FIDO models. In the {top and bottom
panels of the left-hand column of Fig.~\ref{fig19} we plot the
optimized CSUE model} distributions along $\delta$ and $\Delta$
axes, respectively. We see that each of these distributions are much
closer to the {observed distributions} than those shown in
Fig.~\ref{fig14} even though the lowest observed $\delta$ values
cannot be reproduced. The corresponding $g(w_{\alpha_i})$ values are
indicated on each panel of Fig.~\ref{fig19}. However, the
{$w_{\alpha_i}$} sets that minimize the distributions along $\delta$
and $\Delta$ axes (Fig.~\ref{fig19}a,~c) have considerable
differences. In order to compromise these differences we considered
a common $(w_{\alpha_i})$ set that contains the average values of
the two independent original $(w_{\alpha_i})$ sets. The
distributions and the new $g(w_{\alpha_i})$ values corresponding to
these average $(w_{\alpha_i})$ values are shown in the right-hand
panels of {Fig.~\ref{fig19}}. We see that the new distributions
{also differ considerably} from the observed ones. This appreciable
degradation implies an intrinsic limitation of CSUE models in
reproducing the observed statistics and it is not related to the
intrinsic $\alpha$ {probability distribution function} $F_{\alpha}$.

On the other hand, the same analysis for the FIDO models produces
fruitful results. {Figure~\ref{fig20} is similar to
Fig.~\ref{fig19}} but for the FIDO models. The left-hand column
shows that there are $w_{\alpha_i}$ sets that fit {the observed
statistics almost perfectly} along $\delta$ and $\Delta$ axes and
the right-hand column panels imply that these optimizing
$w_{\alpha_i}$ sets are close enough to each other so that their
average values still result in model distributions that are in
perfect agreement with the {\em Fermi} observations. The success of
FIDO models allows the use of the weights $w_{\alpha_i}$ as
estimators of the intrinsic distribution function of $\alpha$ in
observed pulsars. {Figure~\ref{fig21}} shows the {probability
distribution function} $F_{\alpha}(\alpha)$ estimated from the
optimizing weights $w_{\alpha_i}$ used in the right-hand column of
{Fig.~\ref{fig20}}. {We observe that there is a significant
probability excess for low inclination angles
$(\alpha<45^{\circ})$.} For higher inclination angles the
probability decreases even though it seems that there is an
increasing trend for very high $\alpha$. The form of the $\alpha$
distribution function is consistent either to an initial $\alpha$
function predominant in low $\alpha$ values or to an initial
$\alpha$ function that has been modified considerably by the
alignment torques applied on the stellar surface. Finally, we note
that this distribution function represents the distribution of the
existing pulsars and should not be confused with the $\alpha$
distribution of the observed pulsars. The latter distribution
depends also on the probability of observation with respect to
$\alpha$.

\begin{figure*}
  \centering
  \includegraphics[width=12.5cm]{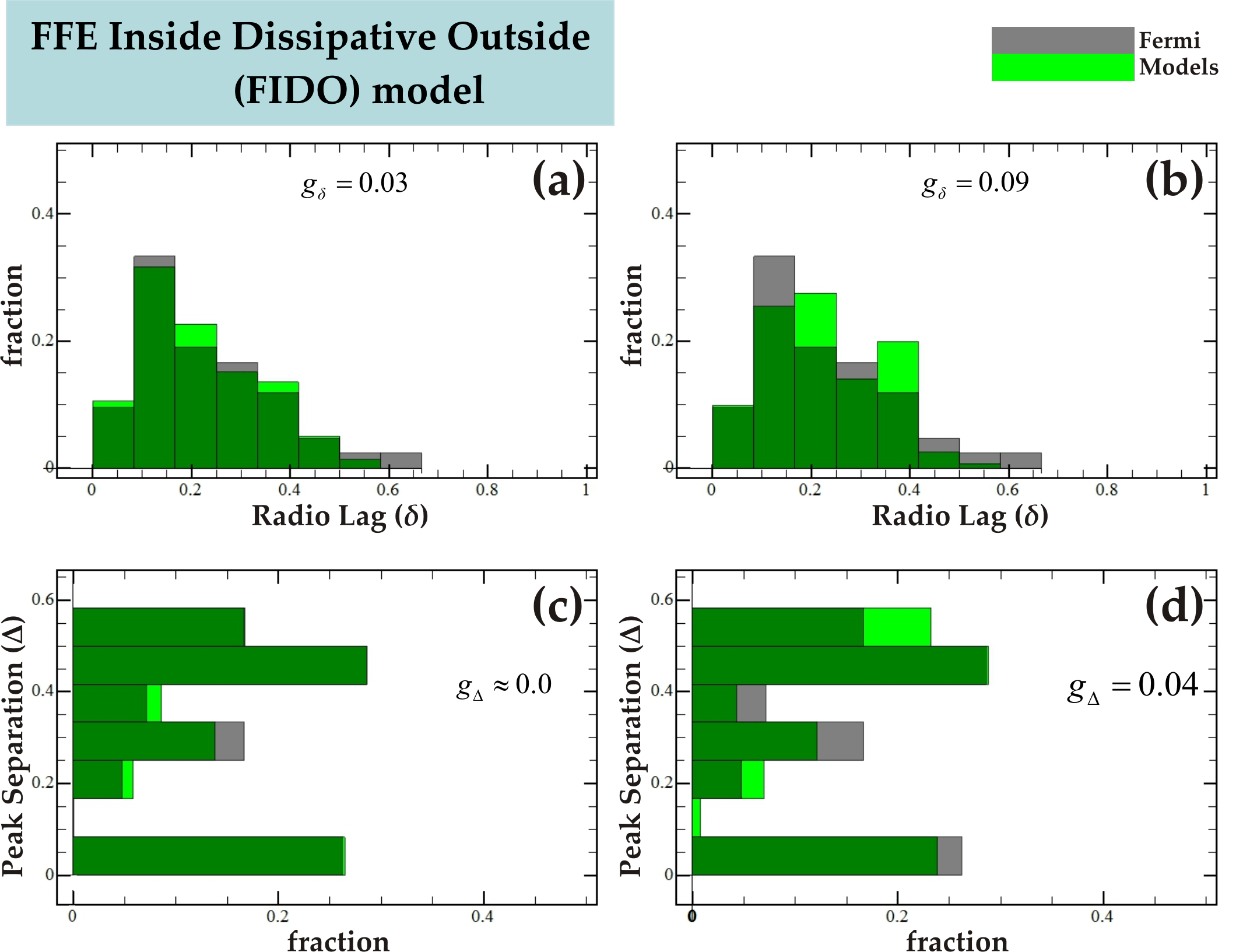}\\
  \caption{Same as Fig.~\ref{fig19} but for the FIDO models. In \textbf{(a), (c)}
  we see an almost perfect match (between the models and the observations)
  for the corresponding optimum $\alpha$ distributions.
  In \textbf{(b), (d)} the mean $\alpha$ distribution of those adopted
  in \textbf{(a), (c)} still produce histograms in good agreement with the
  observations.}\label{fig20}
\end{figure*}

\begin{figure}
  \centering
  \includegraphics[width=7.5cm]{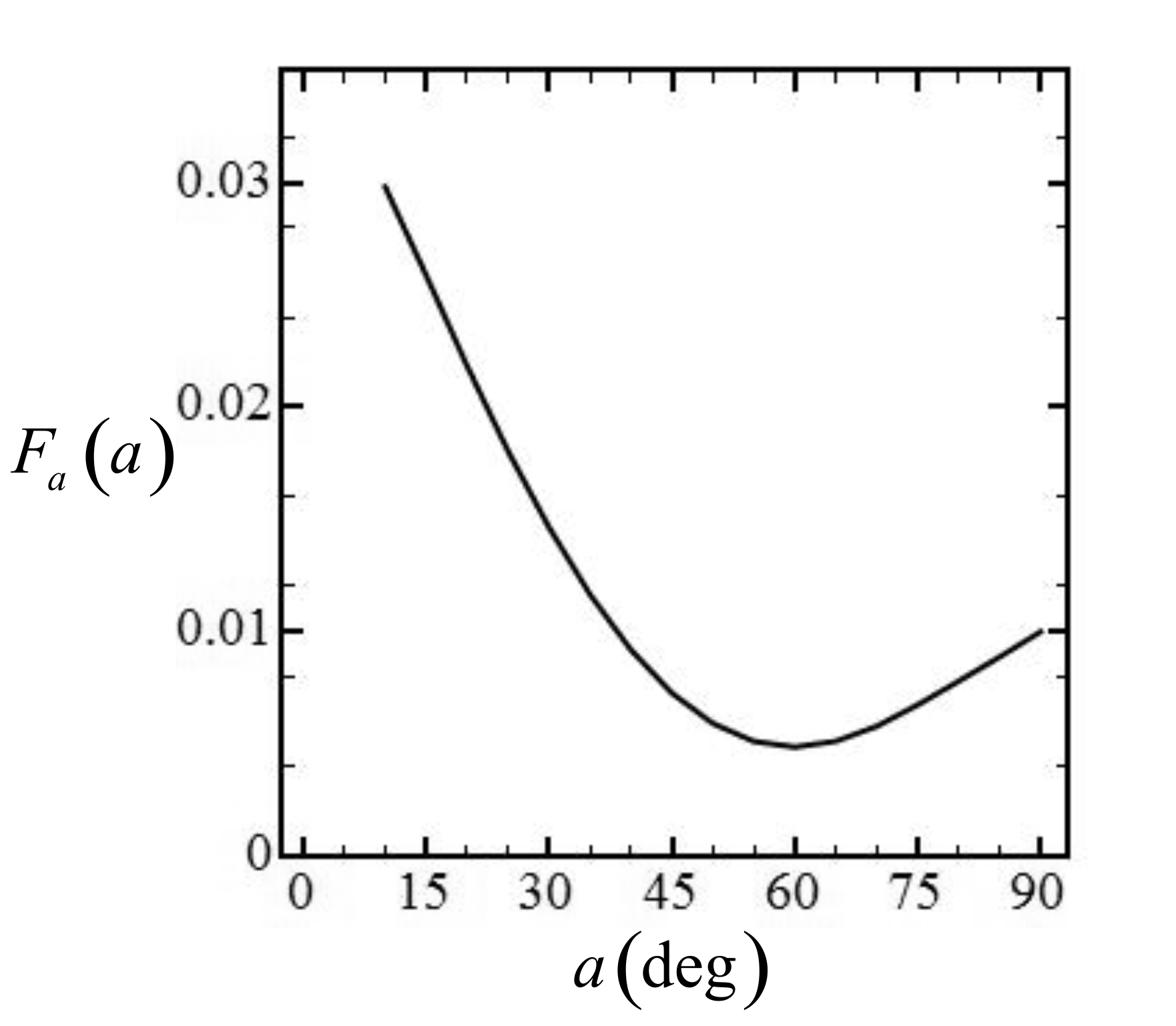}\\
  \caption{The optimum mean $\alpha$ {probability distribution function}
  $F_a(a)$ adopted in Fig.~\ref{fig20}b,~d.}\label{fig21}
\end{figure}

\section{Conclusions}

In this paper we studied the radiation patterns of
$\gamma$-radiation in dissipative pulsar magnetospheres, assuming
the emission is due to curvature radiation. This study enabled us to
identify models that successfully reproduce the observed
$\gamma$-ray light-curve phenomenology.

We considered series of dissipative pulsar magnetosphere models
corresponding to all macrophysical current density prescriptions
used in the literature. These models cover, in general, the entire
spectrum of solutions from the near-vacuum to the near Force-Free
regime and {provide the intrinsic distributions} of parallel
electric field components. These depend on the adopted
conductivity value involved in each dissipative macrophysical
prescription. In general, as the conductivity evolves from low to
intermediate values, both the field geometry and the values of the
parallel electric field components change considerably. However, for
sufficiently large conductivity values ($\sigma \gsim 10 \Omega$)
the geometry stabilizes to near Force-Free structure, but the
accelerating parallel electric components remain sensitive to the
conductivity with their values decreasing with increasing $\sigma$.
For each model, we considered charged particles $(e^+, e^-)$
uniformly distributed on the polar caps and we integrated their
trajectories from the neutron star surface up to a distance 2.5
times that of the light-cylinder radius. We assumed that the
particle velocities consist of two components: (a) a drift component
perpendicular to the plane defined by the electric and magnetic
fields and (b) a component parallel to the magnetic field,
determined by the requirement that the total velocity be $v=c$. The
trajectory integration includes the acceleration of the particles by
the electric fields of the specific model magnetosphere and their
energy losses due to curvature radiation with the local radius of
curvature provided by the model field structure. From this
calculation we produce sky-maps and the corresponding $\gamma$-ray
light-curves based on the physical properties of the models. Our
ultimate goal was to compare the $\gamma$-ray light-curve
phenomenology predicted by the models with those of the observed
pulsars in the Second {\em Fermi} Pulsar Catalog (2PC).

Initially, we considered models {of uniform conductivity} in the
open field line regions (in the closed field line region the
Force-Free condition has been applied i.e. infinite conductivity).
For low conductivity values the radiation is produced mainly in the
inner magnetosphere, well within the light-cylinder. An increase of
conductivity pushes the radiation emission to higher altitude while
the solution gradually starts forming features of the Force-Free
solutions (e.g. current sheets). The formation of an equatorial
current sheet outside the light-cylinder {that extends inside the
light-cylinder along a part of the separatrix} induces high parallel
electric components in the nearby regions. For high conductivity
values {($10\Omega < \sigma < 30\Omega$)} a significant part of the
total radiation is produced in regions near the equatorial current
sheet outside the light-cylinder and for very high values ($\sigma
\gg 30\Omega$) all radiation originates in this region. However, the
emission is not uniform in this region near the equatorial current
sheet since it is modulated by the inclination-dependent
distribution of the accelerating parallel electric field components
{within the light-cylinder}. {The emission in these equatorial CS
regions is more efficient the higher the Lorentz values $\gamma_L$
of the radiating particles. We found that the highest values of
$\gamma_L$ are those of the particles that encounter the highest
parallel electric fields within the LC. These highest field regions
are the segments of the separatrix along which the return current
flows, in current sheet form, down to the stellar surface. This
region is uniform over the separatrix only for the aligned rotator
but as the inclination angle increases the current sheet region of
the separatrix decreases and vanishes for the limiting case of the
orthogonal rotator (see Fig.~\ref{fig01}). This effect makes the
emission in the region near the equatorial current sheet non-uniform
especially for the high inclination angles.

The fitting of model light curves to the observed light curves picks
out the best parts of parameter space for each model, but does not
statistically test the ability of the model phase space to match the
entire distribution of light curve properties.  Instead, we
statistically compared the distribution of the radio-lag ($\delta$)
vs peak-separation ($\Delta$) of the  model and observed light
curves, assuming (for the model light-curves) that the radio
emission is produced near the magnetic poles, not far above the
stellar surface {(compared to the LC radius)}.

This comparison revealed that the low conductivity models give the
poorest match to the  observed $\delta$ vs $\Delta$ distribution,
giving the largest radio-lags of all the models. The higher
conductivity models perform better, as the emission moves outward
and the regions near the equatorial current sheet become gradually
active.  However, even for very high conductivity values there are
still model points on the $(\delta-\Delta)$ diagram that do not lie
near the observed ones. We note also that for these high
conductivities the $\gamma_L$ values of the particles in the
emitting regions do not exceed $\simeq 10^6$ and the corresponding
photon energies are well below GeV. Nevertheless, this analysis
clearly showed that the high conductivity models did best at
matching the data. Moreover, this study identified the different
emission regions near the equatorial current sheet.

The equatorial current sheet had already been proposed as a
candidate source region of the $\gamma$-ray emission
\citep{ck2010,bs10b,2011MNRAS.412.1870P,2013A&A...550A.101A}. These
studies assumed uniform (azimuthally symmetric) emissivity or
constant $\gamma_L$ values over the emitting region. Although we
demonstrated the systematically higher radio-lags of the
light-curves of these models, they seem to produce the observed
trend in the $\delta-\Delta$ diagram, and are statistically better
than the high conductivity dissipative models discussed above.

The relatively good performance of the models that assume ad hoc
uniform emission near the equatorial current sheet and of the high,
uniform conductivity dissipative models, that produce emission in
regions near the equatorial current sheet that is modulated by the
accelerating electric field components in the inner magnetosphere
(within the light-cylinder), motivated us to search for models that
produce most of the $\gamma$-ray emission from regions near the
equatorial current sheet, but this time modulated by the local
physical properties.  We found a simple macrophysical model that
statistically best reproduces the $\gamma$-ray light-curve
phenomenology.

Such a model assumes that the magnetosphere is in an exact
Force-Free regime inside the light-cylinder and dissipative outside
(FIDO) the light-cylinder. This implies that the conductivity in the
inner magnetosphere is infinite while in the outer magnetosphere it
is finite, though still high. The $\gamma$-ray emission is still
produced in a region near the equatorial current sheet but the
emission distribution is different from that of the uniform
conductivity models discussed above. The success of this kind of
model consists mainly in that: \textbf{(a)} The $\gamma$-ray
emission is not present for low observer angle values $(\zeta)$ and
high inclination angles $(\alpha)$ due to the smaller accelerating
electric field in the corresponding regions. The emission at these
angles would come from a region that is relatively close to the
rotational equator, eliminating some points of the $(\delta-\Delta)$
diagram of higher radio-lag values $(\delta)$. \textbf{(b)} The
$\gamma$-ray emission is asymmetric across the equatorial current
sheet, being more effective on the side that comes from the leading
part of the polar cap; this side produces lower radio-lag values
$(\delta)$.

{We note also that the light curves we get in FIDO models for
$\sigma=30\Omega$ changes only slightly for much higher conductivity
values ($\lesssim 1000\Omega$). However, the magnitude of the finite
conductivity value $(\sim30 \Omega)$ of FIDO models, presented in
Section~5, seems to be important.} {Much lower conductivity values
$(\ll 30\Omega)$ are expected to destroy the equatorial current
sheet and the associated high accelerating electric field
components.} {Moreover, the value of conductivity of FIDO models
produces accelerating electric field components and associated
voltages that are able to produce Lorentz factors $(\gamma_L)$ up to
the order of $10^{7.5}$.} With these $\gamma_L$ values the
associated photon energies can reach up to GeVs. {Much higher
conductivity values $(\gg 30\Omega)$ result in a photon energy that
barely exceeds MeVs even though they keep a similar emission pattern
(with narrower pulses).} Thus, the FIDO value seems to support both
the observed light-curve phenomenology and the observed photon
cut-off energies. We constructed the FIDO model of $\gamma$-ray
emission with the goal of reproducing the observed $\gamma$-ray
light-curve phenomenology, and its success in statistically best
matching the observations is important in itself. Now, the fact that
the model emitting regions that emerge naturally as a result of
physical properties of the solutions also accelerate particles to
$\gamma$-ray emitting energies makes them even more significant.

The FIDO emission geometry and sky-maps, while distinctly different
from any of the previously proposed $\gamma$-ray emission models, do
have elements of each. Slot-gap (SG) models \citep{mushar2004} have
a two-pole caustic \citep{dykrud2003} emission geometry, produced by
uniform emission along the closed/open field boundary of both
magnetic poles from the neutron star surface to the light-cylinder.
An observer thus sees emission from both magnetic poles and there is
emission throughout the sky-map and at all pulse phases. Outer-gap
(OG) models \citep{romyad1995} have a one-pole caustic geometry,
produced by emission along the closed/open field boundary but only
in the outer magnetosphere above the null charge surface. An
observer thus sees emission from only one magnetic pole and much of
the sky-map is devoid of emission for smaller inclination angles,
with no emission at off-peak pulse phases. The FIDO model has
intrinsically a two-pole geometry, like the SG model, but no
emission at low altitudes or at off-peak phases, like the OG model.
Although the location of emission in the FIDO model is similar to
that of the current sheet models, the emission pattern {(for high
$\alpha$ values)} looks quite different, with emission over much
less of the sky-map.

The FIDO models imply that the parallel electric field is
efficiently screened in the inner magnetosphere, thus prohibiting
any modulation from the inner magnetosphere of the emission in
regions near the equatorial current sheet. This implies in turn that
pair cascades must be operating along all open field lines;
something that seems to be supported by recent studies of
time-dependent pair cascades
\citep{2010MNRAS.408.2092T,2013MNRAS.429...20T,2013ApJ...762...76C}.
An explanation of why these cascades are not efficient enough to
screen the accelerating electric fields in the outer magnetosphere
could be the following: on the one hand, the large number density of
the charges produced above the polar caps decreases with distance
from the stellar surface due to the diverging flow lines. On the
other hand, the corresponding Goldreich-Julian density could either
decrease slower than the number density or increase with the
distance from the stellar surface (especially for the flows that
pass through regions near the equatorial current sheet; see eq.~6
and eq.~53 in \citealt{ckf99} and \citealt{timokhin2006},
respectively). This effect generally increases the number of
particles required for efficient screening of the accelerating
electric fields. This increased demand on screening arises rather
abruptly near the light-cylinder, where the required
Goldreich-Julian density is high (in regions near the equatorial
current sheet). Even though the demand for charges increases, the
ability of new particle creation is small due to the significantly
smaller magnetic fields and the low probability of photon-photon
interactions.

We note that the theoretical consideration discussed above remain
speculative as long as its validity is not verified by
self-consistent studies of the microphysical processes.  Certainly
this will act as a driver for these kind of studies. The
$\gamma$-ray luminosities produced by the FIDO models should also be
investigated. The microphysical description presented in this paper
allows the derivation of spectra (averaged and phase-resolved) that
can be compared directly with the {\em Fermi} observations. We plan
to present the study of the energetics of FIDO models in a
forthcoming paper. We expect this study to put stronger constraints
on the models and to help in the deeper understanding of the
underlying microphysics.

\bibliographystyle{apj}
\bibliography{references_all}

\end{document}